\documentclass[english,12pt]{article}
\usepackage{scicite}
\usepackage{mathtools}
\usepackage{times}
\usepackage{lineno}
\usepackage{siunitx}
\usepackage{titling}
\usepackage{color}
\usepackage{amsmath}
\usepackage{amssymb}
\usepackage{graphicx}

\usepackage{changes}
\usepackage[unicode=true]
 {hyperref}
\usepackage{hyperref}
\hypersetup{
    colorlinks=true,
    linkcolor=blue,
    filecolor=magenta,      
    urlcolor=blue,
    citecolor=blue,
}
\usepackage[nameinlink]{cleveref}
\Crefname{equation}{equation}{equations}

\usepackage{comment}

\usepackage{tabularx}
\newcolumntype{L}[1]{>{\raggedright\arraybackslash}p{#1}} 
\newcolumntype{C}[1]{>{\centering\arraybackslash}p{#1}} 
\newcolumntype{R}[1]{>{\raggedleft\arraybackslash}p{#1}} 

\newcommand{\rmax}{{r_{\mathrm{max}}}}
\renewcommand{\figurename}{\textbf{Figure}}

\begin{document}
\title{Predictive power of testing for inferring COVID-19 epidemic dynamics and implications for public policy.}
\title{Can testing for SARS-CoV-2 be excessive? A Data-based evaluation of large-scale testing models.}
\title{How to interpret SARS-CoV-2 numbers? A data-based evaluation of large-scale testing models.}
\title{How to Interpret COVID-19 Case Numbers?\\ Accurate large-scale testing models are key to capture  epidemic dynamics.}
\title{Modeling Large-scale Testing is Key to Interpret Covid-19 Case Numbers. A Cross-country Study.}
\title{Assaying Large-scale Testing Models to Interpret COVID-19 Case Numbers.}

\author{Michel Besserve$^{1,2,*}$, Simon Buchholz$^1$ and Bernhard Sch\"olkopf$^1$}
\date{\small $^1$ Max Planck Institute for Intelligent Systems, T\"ubingen, Germany.\\
$^2$ Max Planck Institute for Biological Cybernetics, T\"ubingen, Germany. \\
$^*$ Correspondence to: michel.besserve@tuebingen.mpg.de
}
\maketitle

\baselineskip24pt

\begin{abstract}
\linespread{1.6}\selectfont
Large-scale testing is considered key to assess the state of the current COVID-19 pandemic. Yet, the link between the reported case numbers and the true state of the pandemic remains elusive. We develop mathematical models based on competing hypotheses regarding this link, thereby providing different prevalence estimates based on case numbers, and validate them by predicting SARS-CoV-2-attributed death rate trajectories. 
Assuming that individuals were tested based solely on a predefined risk of being infectious implies the absolute case numbers reflect the prevalence, but turned out to be a poor predictor, consistently overestimating growth rates at the beginning of two COVID-19 epidemic waves. 
In contrast, assuming that testing capacity is fully exploited performs better. This leads to using the percent-positive rate as a more robust indicator of epidemic dynamics, however we find it is subject to a saturation phenomenon that needs to be accounted for as the number of tests becomes larger.

\end{abstract}


\section*{Main}
Assessing the spread of infectious diseases such as the novel coronavirus disease 2019 (COVID-19) epidemic caused by SARS-CoV-2 is a major challenge for modern societies. Reported case number trajectories reflect in principle the epidemic dynamics, and are thus used within the scientific community to infer its evolution \cite{Giordano2020, Dehningeabb9789}, 
but are also publicly reported and debated 
\cite{hart2020mediacoverage, Tejedor2020, Cinelli2020}, 
along with the effectiveness of governmental interventions in lowering them. 
However, the number of reported cases may be influenced by multiple factors, notably depending on the public policy regarding testing, such that its relationship to epidemic dynamics needs to be clarified.

\begin{figure}
	\vspace*{-1cm}
	\hspace*{-2cm}\includegraphics[width=1.2\textwidth]{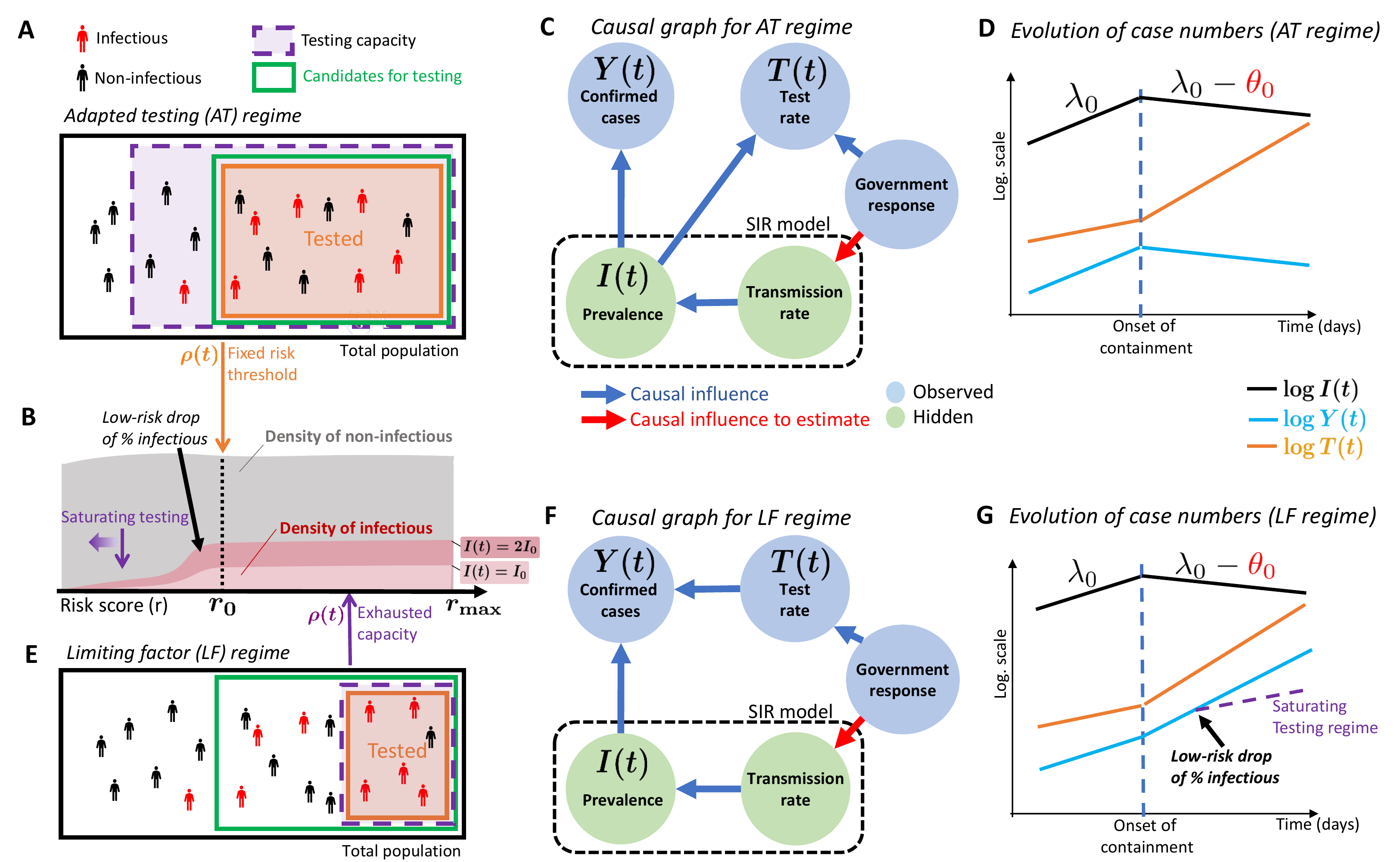}
	\caption{
		\textbf{Different testing regimes in a risk-structured population model.} 
		\textbf{(A)} Schematic of the AT regime, relying on a risk-structured population model described in B, in which all and only individuals with a risk score above $r_0$ are tested (orange rectangle) while testing capacity is larger (violet rectangle). 
		\textbf{(B)} Densities of infectious and non-infectious individuals as a function of risk score. The density of infectious is represented for a given prevalence $I_0$ in light pink, to which the dark pink region is added for a doubling of the prevalence. The gray and pink areas from $r_0$ to $r_{max}$ reflect the number of negative and positive test outcomes for panel A, respectively. A drop in the proportion of infectious at low risk scores affects the evolution of case numbers in the saturating testing regime described in G (black and violet arrow). 
		\textbf{(C)} Causal graph between observed (blue) and hidden (green) variables in the AT regime. The red arrow represents the causal effect of measures on the epidemic dynamics we wish to evaluate.  
		\textbf{(D)} Schematic time course of prevalence (black), number of tests (orange) and of confirmed cases (cyan), for the AT regime. The dashed line indicates the onset of containment measures, inflecting the initial growth rate $\lambda_0$ of the prevalence. 
		\textbf{(E)} Same as A for the LF regime, in which limited testing capacity constrains the number of tested individuals (violet arrow) to a minimum risk score $\rho(t)$ above the threshold $r_0$ set for candidates (green rectangle).
		\textbf{(F)} Same as C for the LF regime, with the number of tests now influencing case numbers. 
		\textbf{(G)} Same as D for the LF regime. The violet dashed line indicates the drop in the growth rate of case numbers, resulting from saturating testing (see B).
		\label{fig:theory}}
\end{figure}

We first investigate this question theoretically with a SIR model of the epidemic, and characterize the dynamics at a given time $t$ by the instantaneous exponential growth rate $\lambda(t)$ of the prevalence $I(t)$, i.e. the total number of infectious individuals at that time \cite{kermack1932contributions,kermack1933contributions} (see \nameref{ssec:matmeth}, section \nameref{ssec:SIR}). In the initial phase of an epidemic wave, $\lambda(t)$ is assumed constant and equals $\lambda_0$, reflecting the baseline disease transmission rate before the onset of governmental response. This implies that prevalence evolves linearly in logarithmic scale with slope $\lambda_0$, as illustrated in Fig.~\ref{fig:theory}D. After the onset of containment measures, the growth rate is reduced to a lower value $\lambda_0-\theta_0$ such that the log-prevalence overall evolves as a piecewise linear function of time, and the absolute change in slope, $\theta_0$, quantifies the causal effect of containment measures on the transmission rate (see Fig.~\ref{fig:theory}D and \nameref{ssec:matmeth}, section \nameref{ssec:SIR}).

To model how this change is reflected in the number of confirmed cases depending on testing policy, we use similar abstraction principles as for continuous age-structured population models \cite{m1925applications} and define a \textit{risk-structured} population model as follows. The idealized large-scale testing mechanism relies on a \textit{risk score} $r$ attributed to each individual. This risk score accounts for observable factors officially used by governments to determine an order of priority for testing (e.g. symptoms, contacts with positive cases, ...), as well as for unobserved factors influencing the likelihood of a given individual to be tested (e.g. location specific limitation of testing capacity). At any time the population being actually tested is assumed to be comprised of \textit{all and only} the individuals above a given minimum risk score $\rho(t)$.
Given this framework, three paradigmatic testing regimes are considered in the following paragraphs. 

First, in the \textit{adapted testing} (AT) regime, tests are performed for all individuals with risk above a constant risk threshold $r_0$ imposed by government policy, such that $\rho(t) = r_0$ at all times $t$. This entails testing capacity allows to handle the growing pools of candidates set by the policy as the epidemic spreads. The rate of cases thus remains uninfluenced by the number of tests performed, which only \textit{adapts} to the number of candidates. To represent the mechanisms leading to the observed case and test numbers, we use causal graphs, which allow evaluating causal effects and learning mechanisms from data in a principled way \cite{causality,elementsofcausal}. Influences between variables for the AT regime are summarized by the causal graph of Fig.~\ref{fig:theory}C, showing that the (unobserved) true prevalence is the only direct cause of the observed case numbers $Y(t)$. This allows to infer the causal effect of government response on prevalence from the time course of $Y(t)$  using the above mentioned SIR model. Indeed, under simplifying assumptions (see \nameref{ssec:matmeth} section~\nameref{ssec:testModels}), the total number of positive cases is then  proportional to the prevalence of the disease, $I(t)$, which modulates linearly the risk-structured density of infectious individuals illustrated in Fig.~\ref{fig:theory}B. Prevalence variations can thus be estimated based on case numbers variation, irrespective of the number of tests performed $T(t)$, as illustrated in Fig.~\ref{fig:theory}D.

One limitation of the AT regime is that the demand for tests does not necessarily match the supply. We thus introduce a second regime modeling a limited testing capacity, which we assume fully exploited to perform the maximum number of tests while prioritizing larger risk scores. This \textit{limiting factor} (LF) regime notably models the inability to test all candidates indicated by the government policy threshold $r_0$. The number of tests performed, reflecting testing capacity, then becomes a limiting quantity directly influencing the rate of confirmed cases $Y(t)$ as depicted in Fig.~\ref{fig:theory}E and the corresponding causal graph of Fig.~\ref{fig:theory}F.  When estimating the effect of public policy on epidemic dynamics, this graph configuration indicates a caveat called \textit{confounding}: government response affects simultaneously the number of tests and the containment policy, thereby influencing case numbers along two different causal pathways. 
As a consequence, the modification of the case numbers' time course are not solely imputable to containment measures on transmission rate, but also to concurrent effects of policy on testing capacity. Not accounting for changes in the number of tests may thus lead to misleading interpretations of epidemic dynamics based on the case number time course. 
A putative example is depicted in Fig.~\ref{fig:theory}G, where the growth rate of the number of tests increases concomitantly with a switch from growth to decay of prevalence due to social distancing measures instated at the onset of lockdown. Under simplifying assumptions (see \nameref{ssec:matmeth}, section~\nameref{ssec:testModels}), reported cases then become proportional to the number of tests actually performed, reflecting the naive intuition: ``the more we test, the more we find cases''. The case numbers are then approximately given by
\begin{equation}\label{eq:massAction}
Y(t) \approx \kappa I(t) T(t)
\end{equation}
where $T(t)$ is the test rate at time $t$ and $\kappa$ is a multiplicative constant.  This relation leads to a growing number of cases in Fig.~\ref{fig:theory}G, that clearly misrepresents the true epidemic dynamics (characterized by the prevalence) and the effect containment measures have on it.

Finally, the above limiting factor regime can evolve into a \textit{saturating testing} (ST) regime as the testing capacity increases during the course of the epidemic, allowing to satisfy the demand for testing individuals with smaller and smaller risk scores. 
Going below a certain risk level may result in an drop of the probability of a single test to detect a true positive (see Fig.~\ref{fig:theory}B). For example, generalizing testing to asymptomatic individuals is likely to reduce the probability of detecting positive cases among them. In the example of Fig.~\ref{fig:theory}G (see violet dashed line), this leads to a milder growth rate of the number of cases compared to the prediction by the LF regime model of equation~\ref{eq:massAction}, again misrepresenting the epidemic dynamics if not accounted for.

\newcommand{\Nworldvalid}{66}
\newcommand{\Nworldvalidcovariates}{62}
\newcommand{\NvalidUS}{40}
\newcommand{\pcompareBasePoisson}{\SI{3.0e-4}{}}
\newcommand{\pcompareDownPoisson}{\SI{1.0e-2}{}}
\newcommand{\pcompareUpPoisson}{\SI{2.1e-3}{}}

\newcommand{\pcompareBasePoissonUS}{\SI{2.6e-7}{}}
\newcommand{\pcompareDownPoissonUS}{\SI{2.8e-4}{}}
\newcommand{\pcompareUpPoissonUS}{\SI{7.4e-4}{}}

\begin{figure}
	\vspace*{-2.8cm}	\hspace*{-.5cm}	
	\includegraphics[width=.9\textwidth]{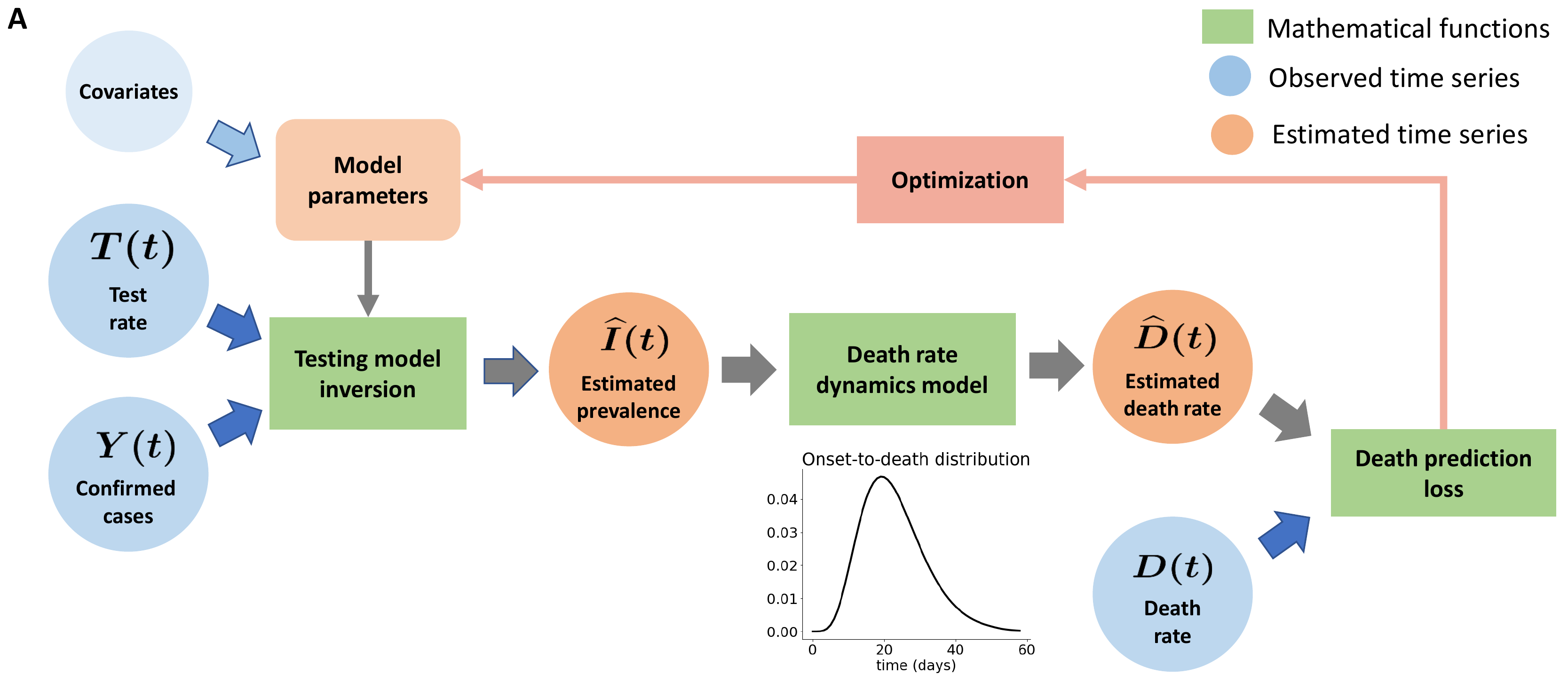}
	
	\hspace*{-.7cm}
	\includegraphics[width=1.0\textwidth]{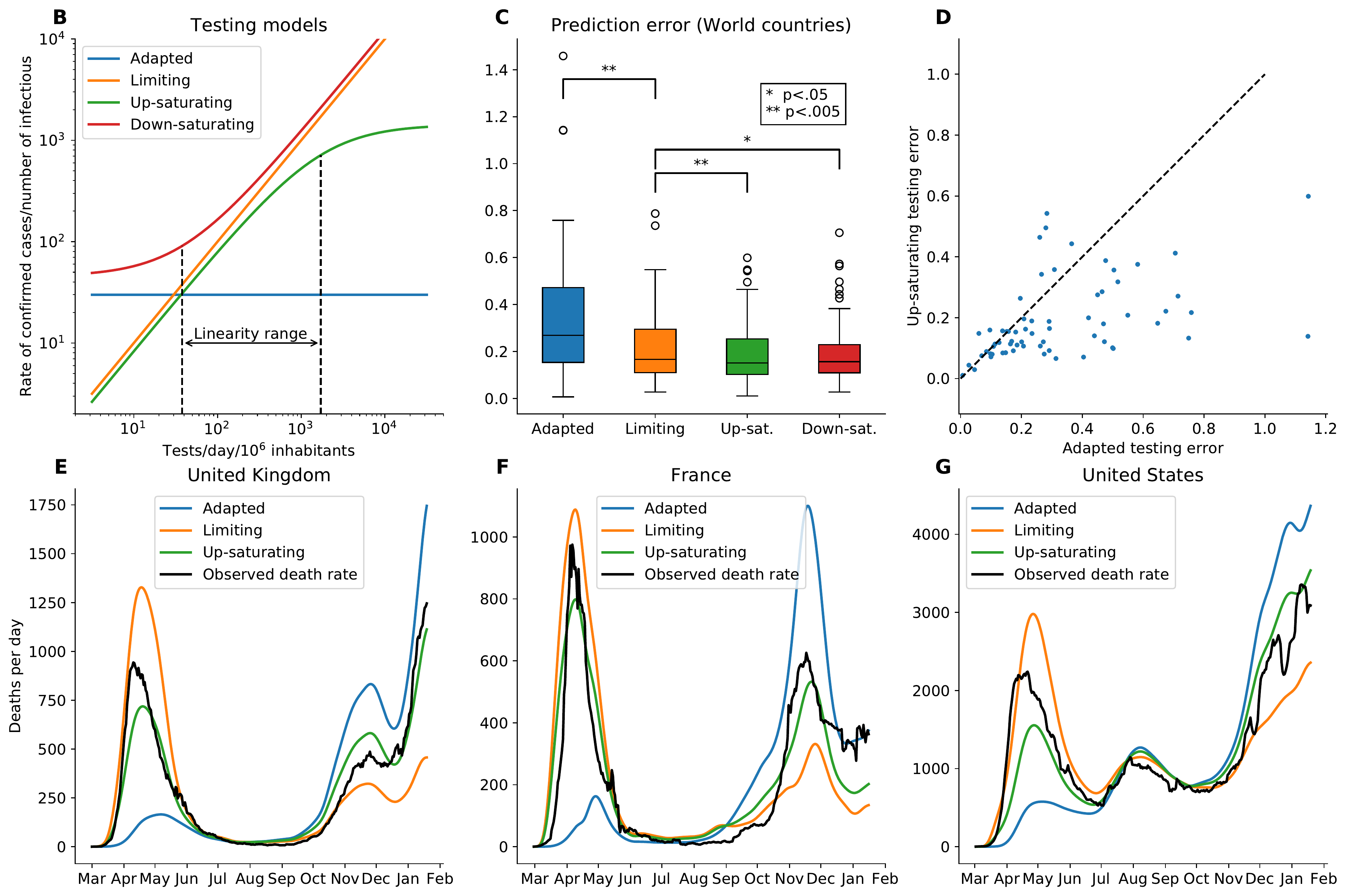}
	\caption{\textbf{Validation of population testing models based on death rate prediction.} \textbf{(A)} Schematic of the computational pipeline for death rate prediction based on observed time-series, mapping testing numbers to inferred death rate, and optimizing the testing model parameters to minimize the prediction error of the true death rate. Country covariates are allowed to modulate the model parameters. \textbf{(B)} Functions associated to different testing models, with median parameters fitted by the procedure in A (without covariates) on the World countries dataset. Dashed line indicate upper and lower saturation threshold parameters, indicating the \textit{linearity range} of daily tests per inhabitant for which saturating models departs from the limiting factor regime by less than a factor two. \textbf{(C)} Country-wise cross-validation prediction error for  death rate using different models linking these quantities to incidence. P-value is obtained by paired Wilcoxon signed-rank test. \textbf{(D)} Scatter plot showing the superiority of the limiting factor testing model, relative to the adapted testing model. \textbf{(E-G)} Prediction of the time course of the empirical death rate (in black) for 3 example countries, based on inferring the number of infectious individuals, either using the limiting factor (in orange), the adapted testing model (in blue) and the up-saturating model (in green). See also Supplementary Figs.~\ref{sfig:modelval}-\ref{sfig:covariates}.
		\label{fig:validModel}}
\end{figure}

We next investigate which of the above large-scale testing regimes describe best the current COVID-19 epidemic based on empirical data. 
We evaluate the validity of models associated to each regime based on their ability to predict the evolution of death rates in two datasets: the \textit{World Countries} testing dataset of \textit{Our World in Data}\footnote{available at \href{ourworldindata.org/coronavirus}{ourworldindata.org/coronavirus}}
\cite{Hasell2020} containing time resolved evaluations of the number of tests and confirmed cases for many countries across the world, and the \textit{US States} dataset from the COVID Tracking Project 
\footnote{available at \href{covidtracking.com}{covidtracking.com}} that contains similar data for all US states. As described in the schematic of Fig.~\ref{fig:validModel}A, the testing models provide an estimate of the prevalence which can be exploited to predict the variations of deaths rate $D(t)$ across time using a convolution with the onset-to-death distribution, describing the stochastic latency between the time of infection and time of death in the population, and which has been measured experimentally and subsequently modeled \cite{verity2020estimates,flaxmanNature} (see \nameref{ssec:matmeth}, section \nameref{ssec:infectToDeaths}).

The testing models include the regimes described in Figs.~\ref{fig:theory}A and \ref{fig:theory}E: on the one hand the \textit{adapted testing} model is our baseline, in which the rate of cases $Y(t)$ reflects $I(t)$ up to a multiplicative factor; on the other hand, the limiting factor model implies that the rate of cases is proportional to the product of $I(t)$ and $T(t)$ (based on equation~(\ref{eq:massAction})). We additionally included two models accounting for a possible saturation of the limiting factor regime, as the number of tests becomes too large (the \textit{up-saturating} testing regime described in Fig.~\ref{fig:theory}B and G) or as the number of tests becomes too small (the \textit{down-saturating} regime, see \nameref{ssec:matmeth}, section \nameref{ssec:testModels}). 
All models are associated with a \textit{testing function} $f_m$ (where $m$ indicates the model), parameterized by one scalar coefficient in the case of both saturated models, and linking the number of confirmed cases to prevalence through
\begin{equation}\label{eq:testingModelGeneric}
Y(t) = I(t) f_m(T(t))\,.
\end{equation}
Each model is thus naturally associated to an estimate of prevalence $\widehat{I}(t)$ such that
\begin{equation}\label{eq:prevalenceEst}
	\widehat{I}(t) = \frac{Y(t)}{   f_m(T(t))}\,.
\end{equation}
This framework is exploited to compare testing models based on
and their ensuing ability to predict the observed death rate $D(t)$, with estimate $\widehat{D}(t)$. For that we perform cross-validation of these models across countries, optimizing for putative free parameters of each model (see \nameref{ssec:matmeth} section \nameref{ssec:validTest}). Optimization of the parameters was implemented using the automated differentiation capabilities of the pyTorch library \cite{NEURIPS2019_9015} and the BFGS algorithm (see \nameref{ssec:matmeth} section~\nameref{ssec:optim}). We estimated the prediction error of the logarithmic death rate of each model using 10-fold cross-validation across 
$ \Nworldvalid$ countries (see \nameref{ssec:matmeth}, section \nameref{ssec:excl_sel} for selection criteria).

The functions associated to different testing models are shown (up to a multiplicative constant) on Fig.~\ref{fig:validModel}B (see \nameref{ssec:matmeth}  section~\nameref{ssec:testModels} for their theoretical justification), using the median parameters fitted on the World Countries dataset for the Up- and Down-saturating models. In line with our previous explanations, the number of confirmed cases is independent of the number of tests performed per inhabitants in the adapted model, while there is a linear dependency between these quantities in the limiting factor model. The up- and down-saturating models are respectively characterized by upper and lower thresholds parameters at which the confirmed cases prediction depart from the (linear) limiting factor model by a factor 2. 
Cross-validation results leads to an estimate of the lower threshold of 
38(32,43)
daily tests per million inhabitants and an upper threshold of 
1.71(1.53,2.06)
thousands daily tests per million inhabitants (estimates are median of the values obtained during cross-validation, CI computed using the \textit{MedianCI} function of the \textit{DescTools} R package). 

We further compared cross-validation errors of each model (yielding one error value per country) using paired two-sided Wilcoxon signed-rank tests. 
Figure~\ref{fig:validModel}C shows that the limiting factor model performs significantly better than the adapted testing model
  ($p<\pcompareBasePoisson$, $N = \Nworldvalid$), as further supported by error values of individual countries shown in the scatter plot of Fig.~\ref{fig:validModel}D.
Moreover, taking into account a form of saturation further significantly improved the prediction over the simple limiting factor model. In particular, the up-saturating model shows the strongest improvement ($p=\pcompareUpPoisson$, $N= \Nworldvalid$). This suggests that in the last months, the increase in the number of tests has led to testing individuals with a lower probability of being infectious. This interpretation is in line with the under estimation of death rate by the limiting factor model (corresponding to the percent-positive rate, up to a multiplicative constant) during this period, illustrated on several countries in Fig.~\ref{fig:validModel}E-G (orange plots). Overall, the improvements brought by saturating models suggest the range of 
[38, 1700]
tests per day per million inhabitants provided by their respective thresholds provide a good indication of the domain of validity of the linear behavior of the limiting factor model.

In order to assay testing models at a different spatial scale, we performed the same analysis across $\NvalidUS$ states of the USA (see \nameref{ssec:matmeth}, section \nameref{ssec:usselect}). Again, while the limiting factor model still outperformed the adapted testing model ($p=\pcompareBasePoissonUS$, $N=\NvalidUS$), both up- and down-saturated parametric models further reduced significantly the prediction error ($p=\pcompareUpPoissonUS$ and $p=\pcompareDownPoissonUS$, respectively, for the comparison to the limiting factor model; see Supplementary Fig.~\ref{sfig:modelval}). This supports again the existence of saturation phenomena that make the testing model depart from the purely linear limiting factor regime, with lower threshold 170(160,180) daily tests per million inhabitants, and upper threshold 3.9(3.7,4.3) thousands daily tests per million inhabitants (median value; CI computed using the \textit{MedianCI} function of the \textit{DescTools} R package). 

To improve the predictive power of parametric models on the World's countries dataset, we also investigated ways to handle country heterogeneity. While fitting a separate model to each country is prone to overfitting, we allowed the model to adapt to the country through the influence of country specific covariates on the parameter of the model, as illustrated in Fig.~\ref{fig:validModel}A. We used the country's Gross Domestic Product (GDP)  per capita, population density, and the percentage of urban population as covariates prone to influence the testing capabilities and the risk structure of the population in our model.  Adding these covariates did not consistently improve the model's predictive power 
(see Supplementary Fig.~\ref{sfig:covariates}), suggesting the increase in sample complexity outweighed the benefits of adaptivity to the countries' heterogeneity allowed by this covariate dependency.
 
As shown in the examples of Fig.~\ref{fig:validModel}E-G, using the adapted testing model,
i.e. case numbers,   to infer death rate may lead to severe misevaluation of the timing and magnitude of the death rate peaks, which may lead to inaccurate political sentiment towards the disease and unsuitable political decisions. In contrast, using the limiting factor regime is a major improvement, and has the benefit do be associated to a simple proxy for epidemic dynamics: the percentage of positive tests. However, we have evidence that this quantity turns out to saturate, in particular as more tests are performed, as taking into account this saturation lead to more accurate predictions. Since these results imply that the adapted testing regime inaccurately estimates the prevalence of the disease, we next studied the impact of choosing this assumption for growth rate and policy effect estimations
and compared it to the best performing model according to our analysis: the up-saturating limiting factor model.




\newcommand{\NfirstLockdown}{13}
\newcommand{\NsecondLockdown}{13}

%
\newcommand{\rorigmedianprefirst}{.20\mbox{ days}^{-1}}
\newcommand{\rcorrecmedianprefirst}{.07\mbox{ days}^{-1}}
\newcommand{\rorigmedianprefirstlow}{.16}
\newcommand{\rcorrecmedianprefirstlow}{.00}
\newcommand{\rorigmedianprefirstup}{.26}
\newcommand{\rcorrecmedianprefirstup}{.11}

\newcommand{\prcorrecprefirst}{\SI{2.5e-6}{}}
\newcommand{\prcorrecpresecond}{\SI{1.4e-3}{}}

\newcommand{\porig}{\SI{.064}{}}

\newcommand{\pcorrec}{\SI{8.8e-3}{}}
\newcommand{\peffectorig}{\SI{1.8e-3}{}}
\newcommand{\peffectcorrec}{\SI{1.9e-2}{}}
\newcommand{\peffect}{\SI{1.9e-3}{}}

\newcommand{\peffectorigsecond}{\SI{1.9e-3}{}}
\newcommand{\peffectcorrecsecond}{\SI{1.9e-3}{}}
\newcommand{\peffectsecond}{\SI{1.6e-2}{}}

\newcommand{\pdeltastringency}{\SI{2.4e-4}{}}

\newcommand{\ptestdurfirst}{\SI{4.6e-2}{}}

Using the OWID testing dataset of World countries \cite{Hasell2020}
we investigated epidemic growth rate trajectories around two periods of instatement of stronger social distancing measures, that we call ``lockdown'' for the sake of conciseness, although the additional measures taken are not officially qualified as lockdown in every country. These periods respectively start in the first (called ``first lockdown'') and fourth (called ``second lockdown'') quarters of 2020 in many countries, notably European (see \nameref{ssec:matmeth},  section~\nameref{ssec:lockdown} and Supplementary Tables~\ref{tab:ld}-\ref{tab:ld2} for details).  Missing data were interpolated at the resolution of a single day using cumulative numbers, and rates were estimated using first-order differences between successive days

\begin{figure}
	\vspace*{-3cm}\hspace*{-2cm}\includegraphics[width=1.2\textwidth]{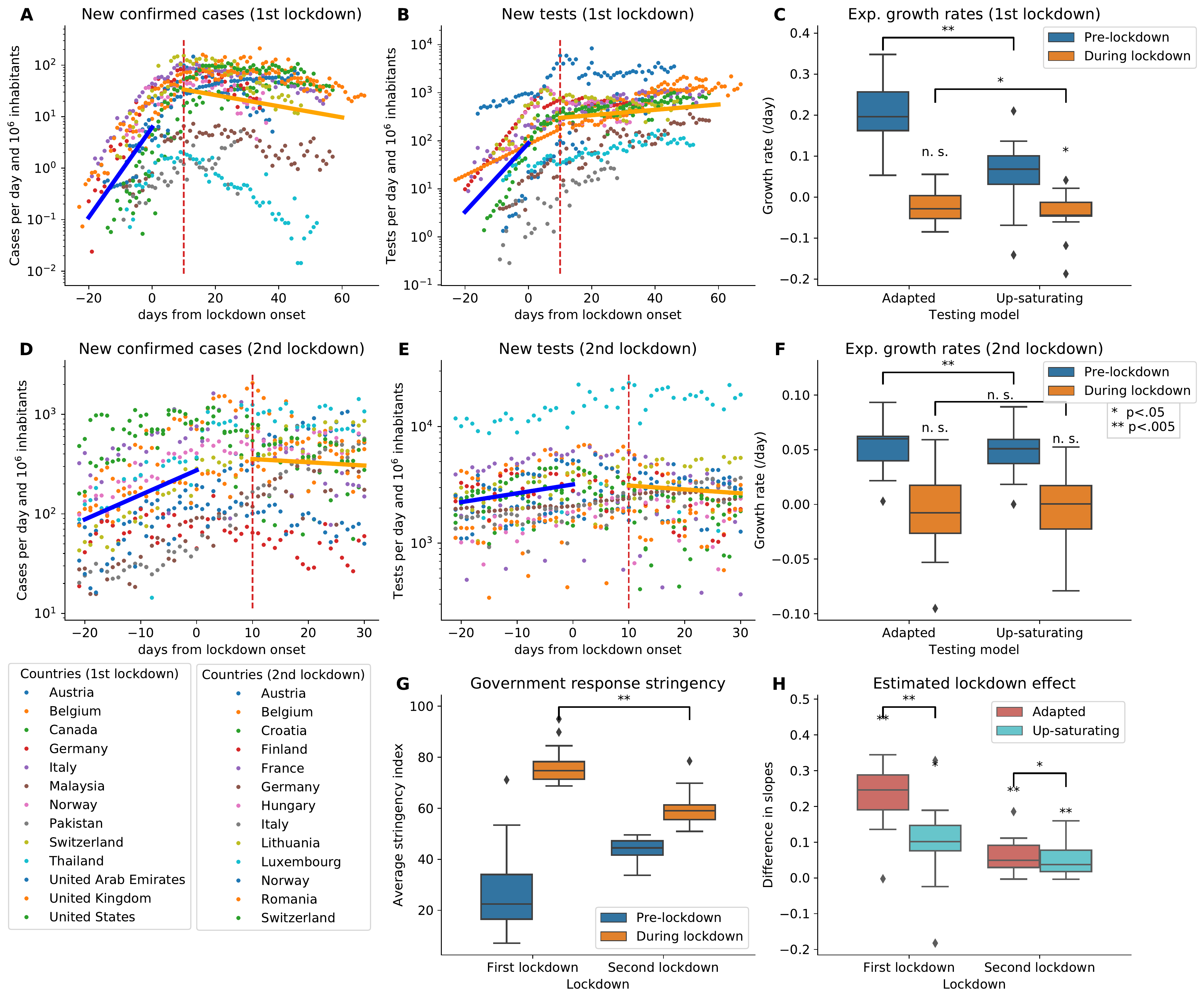}
	\caption{\textbf{Impact of testing assumptions on growth rate estimation and policy evaluation.} \textbf{(A)} Number of daily new confirmed cases for each country for the first lockdown. Blue and orange solid lines represent the average of Poisson regression models across countries, before and during social distancing, respectively. \textbf{(B)} Same as A for daily new tests, first lockdown. \textbf{(C)} Statistics of the exponential growth rate of the  prevalence estimated by Poisson regression for each
		country individually, separately for time periods before and during application of social distancing measures for the first lockdown. Prevalence is estimated according to two testing models: adapted testing (left) and up-saturating (right). \textbf{(D-F)} Same as A-C for second lockdown. \textbf{(G)} Stringency index statistics across countries before and during application of increased measures. \textbf{(H)} Estimate of the effect of social distancing on transmission rate, with or without correction for testing. Indicated p-values correspond to Wilcoxon signed-rank tests on the sign of the median of the regression coefficients or their difference (when above a line connecting two quantities). Country color codes for A-B and D-E are indicated at the figure's bottom left corner. In A-B and D-E, linear patterns at the scale of few days reflect the interpolation of missing data (see \nameref{ssec:matmeth}). See also Supplementary Figs.~\ref{sfig:policyeval}-\ref{sfig:USpolicyeval} and Tables~\ref{tab:ld}-\ref{tab:ld2}.
		\label{fig:policyeval}}
\end{figure}

Figure~\ref{fig:policyeval}A-B shows clear concurrent changes in testing and confirmed cases during the first lockdown.
Interestingly, while the amount of tests increases exponentially before such measures are taken, it tends to plateau after, with variable growth rate signs across countries (see also
Supplementary Fig.~\ref{sfig:policyeval} for testing rates).
We investigate how these variations of testing at the demographic scale impact the estimation of the state of the epidemics. We focus on the exponential growth rate as a marker of this state, as its sign is a key indicator for decision-makers. 
Unless mentioned otherwise, statistical analyses are performed with a (paired, two-sided) Wilcoxon signed-rank test. 
Fig.~\ref{fig:policyeval}C  compares the exponential growth rates for the first lockdown estimated with Poisson regression (see \nameref{ssec:matmeth}) on two time intervals: before and after instatement of more stringent containment measures, and using both adapted and up-saturating testing assumptions. While the estimated growth rate is high for the pre-lockdown period (median: $\rorigmedianprefirst$($\rorigmedianprefirstlow,\,\rorigmedianprefirstup$)) using the adapted testing model, it is significantly smaller 
(median: $\rcorrecmedianprefirst$($\rcorrecmedianprefirstlow,\,\rcorrecmedianprefirstup$))
 when using the limiting factor testing regime assumption 
 ($p=\prcorrecprefirst$; $N=\NfirstLockdown$). This reflects the overestimation of the growth rate before lockdown, due to a concurrent increase in test rates (see Supplementary Fig.~\ref{sfig:policyeval}), which is corrected for by the up-saturating testing model. 
The growth rate estimated during the first lockdown using the adapted model is not significantly negative for the adapted testing model ($p=\porig$; $N=\NfirstLockdown $). 
In contrast, using the up-saturating model leads to a significant negative growth-rate ($p=\pcorrec$; $N=\NfirstLockdown $), as expected, reflecting the effectiveness of the instated measures. This likely reflects the correction of the upward bias of the adapted testing model due to the still slightly increasing test rate during this period ($p=\ptestdurfirst$;$N=\NfirstLockdown$, see Supplementary Fig.~\ref{sfig:policyeval}).
We further evaluated the impact of the testing model assumptions on the estimation of the causal effect of the first lockdown, as measured by the difference between during- and pre-lockdown slopes, computed separately for each country. 
The results on Fig.~\ref{fig:policyeval}H (left-hand boxes) confirm a significant causal effect of containment measures on growth rate for both adapted and up-saturating factor testing models ($p=\peffectorig$ and $p=\peffectcorrec$, respectively; $N=\NfirstLockdown$). However, the up-saturating model leads to a significantly lower estimate of this effect, reflecting the lower value of the initial growth-rate estimate ($p=\peffect$; )$N=\NfirstLockdown$. 

We next ran the same analyses for the second lockdown, in the fourth quarter of 2020 (Fig.~\ref{fig:policyeval}D-F). While the overall initial growth rate prior to lockdown is expected to have moderate magnitude compared to the first lockdown (Fig.~\ref{fig:policyeval}D), likely reflecting the higher initial stringency of measures as shown in Fig.~\ref{fig:policyeval}G, using the up-saturating testing assumption additionally reduces the initial growth-rate
($p=\prcorrecpresecond$, $N=\NsecondLockdown$), suggesting again that the increase of testing prior to lockdown tends to over-estimate the growth of the epidemics at the beginning of this outbreak. In contrast, growth rates during lockdown are not significantly different from zero under both testing assumptions ($p>.05$, $N=\NsecondLockdown$), and of comparable magnitude. This result are in line with the overall unsatisfactory outcome of the containment measures of the second wave, in line with their significantly smaller stringency shown in Fig.~\ref{fig:policyeval}G in comparison to the first wave ($p=\pdeltastringency$, Kruskal-Wallis test; $N=26$).
Finally, the inferred causal effect of the second lockdown, shown in Fig.~\ref{fig:policyeval}H, is significant for both testing regime assumptions, although the up-saturating model again leads to a lower value of this effect in comparison to adapted testing ($p=\peffectsecond$; $N=\NsecondLockdown$). These growth rate and causal effect analyses were additionally replicated at the scale of US states for the first lockdown, supporting again an over-estimation of the pre-lockdown growth rate and of the  causal effect of containment measures (Supplementary Fig.~\ref{sfig:USpolicyeval}).

Overall, we provide evidence that large-scale testing for SARS-CoV-2 is best approximated by an saturated limiting factor regime. This entails that the percent-positive tests can be used as a default statistical proxy for prevalence dynamics instead of the absolute number of confirmed cases, provided the number of daily tests per million inhabitants remains within a range of [38, 1710] ensuring a linear effect on the number of confirmed cases. Above this range, the percent-positive tests progressively become insensitive to prevalence variations, emphasizing the importance of using a non-linear proxy to be able to keep track of the epidemic dynamics. Our current estimate of the this proxy based on death rate prediction is the up-saturating model
\begin{equation}\label{eq:prevalenceEst_estimated}
	\widehat{I}(t) \propto Y(t)\frac{T(t)+1710}{T(t)}\, ,
\end{equation}
with $T(t)$ in daily tests per million inhabitants.

Importantly, other proxies provide a biased perspective on the epidemic dynamics. 
In particular, increases in testing frequency lead to a consistent overestimation of the growth rate at the onset of epidemic waves when relying only on the time evolution of confirmed cases. 
This is likely due to a strengthening of political sentiment towards the disease, concomitant with increased media coverage. The increase in testing rate however tends to plateau after the establishment of stronger social distancing policies. Putative causes for this change may be the achievement of sufficient testing capacity from the perspective of health authorities. An important consequence of this time dependent testing rate is to overestimate the causal effect of the social distancing measures of each epidemic wave and the initial spread of the epidemic, when evaluations are based on the absolute number of confirmed cases. 
Using the serial interval distribution of the pandemic, the growth rate $\lambda$
of an epidemic relates to the reproduction number $R$ \cite{doi:10.1098/rspb.2006.3754}. Our results are thus in line with the observation that estimates for the reproduction number based on observed cases are often higher \cite{Dehningeabb9789, flaxmanNature, 10.1093/jtm/taaa021} than estimates based on the number of fatalities or patients in critical care \cite{Saljeeabc3517}, or cases among flight passengers
	\cite{imai2020report}.	
Therefore, our results suggest that such analyses should either use  more reliable indicators like patients in ICU units or fatalities instead of reported case numbers, or include the most likely testing assumptions in the model. 
Notable works that consider the effect of testing on case numbers include \cite{CHERNOZHUKOV2020, contreras2020longterm} 
and the project \cite{rtlive2020}. However, as these works employ a limiting factor testing model, their estimates are likely to become biased as the number of tests leaves the linearity range, and they would benefit from using the up-saturating model estimated in the present paper.

One limitation of our results is that upon change of testing policy the relevant model may change, and the proposed regimes are simple approximationseq: that can be improved by an accurate and time resolved documentation on the testing process (for example by assessing the evolution of testing capacity). More generally, as more epidemiological data will be gathered, more accurate models of large-scale testing can be estimated. Notably, further characterization of the testing process may be achieved through detailed modeling of the risk score used in different countries and the distribution of infectious and non-infectious individuals according to this score.

Finally, these results suggest that changes in public policy regarding testing must be well-thought-out and documented in order to maintain reliable assumptions on the testing mechanisms, and thus a precise and timely evaluation of epidemic dynamics. Awareness of this aspect of testing at the demographic level are moreover particularly relevant in order to establish optimal epidemic control policies that mitigate economical impacts \cite{karin2020adaptive}.

\section*{Methods}\label{ssec:matmeth}

\subsection*{SIR Model}\label{ssec:SIR}
We use a simplified model of exponential spread of the disease within a country, neglecting the influence of imported cases. We start with the Susceptible-Infectious-Recovered (SIR) model
\begin{align*}
\frac{dS}{dt} & =-\beta(t) S(t)I(t)\\
\frac{dI}{dt} & =\beta(t)S(t)I(t)-\gamma I(t)\\
\frac{dR}{dt} & =\gamma I(t)
\end{align*}
where $\gamma$ is the recovery rate, assumed constant, and $\beta(t)$ is the disease transmission rate which might be influenced by time varying social distancing measures.
 We assume that $I(t)$ (the current number of contagious individuals
in a country) as well as $R(t)$ remain small with respect to the total population, as supported by empirical evidence \cite{bassi2020observed,rostami2020sars},
such that variations of $S(t)$ relative to its initial value remain small.
 This approximation leads to a homogeneous first order differential equation for $I(t)$
\begin{equation}\label{eq:firstOrdI}
\frac{dI}{dt}=\left(\beta(t)S(0)-\gamma\right)I(t)=\lambda(t)I(t)
\end{equation}
where the sign of $\lambda(t)=\beta(t)S(0)-\gamma$ determines whether the outbreak goes on spreading or diminishes. 
The rate $\lambda$ is closely related to the reproduction number $\boldsymbol{R}_0$, which  can be defined as $\boldsymbol{R}_0=\frac{\beta S(0)}{\gamma}$ in the SIR model \cite{van2017reproduction}. 
To infer the epidemic dynamics, we focus on estimating $\lambda(t)$ which is directly related to $I(t)$ as is logarithmic derivative, using equation~\ref{eq:firstOrdI}, such that
\[
\frac{d\log(I)}{dt} (t)= \frac{1}{I(t)}\frac{dI}{dt}(t)=\lambda(t)\,.
\]
Notably, this leads to the exponential form of the prevalence, further exploited in the section \nameref{ssec:causEffect},
\[
I(t)=I(0)\exp\left(\int_{0}^{t}\lambda(u)du\right)\,.
\]

\subsection*{Large scale testing models}\label{ssec:testModels}
\subsubsection*{Risk-based testing framework}
Following the same lines as age-structured population models with continuous age distribution \cite{m1925applications,webb1985theory}, we consider the population as a continuum of individuals, for which we can define densities of individuals satisfying particular properties. 
More specifically, we consider a \textit{risk-structured population} in which
the selection of candidates for testing is done through a random risk variable $R(k)$ that assigns to each individual $k$ a positive risk. Note that for the remainder of the supplementary information, $R$ will always denote the random risk, and not the number of recovered individuals of the above SIR model. The higher the risk, the higher is the priority of the individual for being tested. This risk is assumed continuous valued and reflects not only the observed information (symptoms, contact cases,...) but also exogenous influences that will affect the probability of the individuals for being tested. 

$R(k,t)$ depends on the actual state of the individual $\sigma(k,t)$ (infectious, $\sigma(k,t)=1$, or not, $\sigma(k,t)=0$), and on the overall state of the epidemic ($I(t)$ large makes individuals more likely to be a contact case, thus to have a high risk). As a consequence, using the continuum assumption, $R(k,t)$ is distributed across the population according to a density of the form $p(r|I(t),\sigma(k,t))$ such that the probability of an arbitrary individual in state $\sigma(k)$ to have a risk in the small interval $[r,r+dr]$ is $p(r|I(t),\sigma(k))dr$. 
The average number of infectious individuals in the total population with risk in $[r,r+dr]$
is given by
\[
I_r(t)dr = \mathbb{E}\sum_{k,\sigma(k)=1} 1_{R(k)\in [r,r+dr]} dr= I(t)p(r|I(t),\sigma=1)dr \,,
\]
in the same way, the number of susceptible with a given risk is given by
\[
N_r(t)dr =  N(t) p(r|I(t),\sigma=0)dr\,.
\]

where $N(t)$ is the total number of non-infectious individuals in the above SIR model,
i.e., $N(t)=S(t)+R(t)$. 
It is reasonable to combine the two compartments 
because many infections are not detected so that it is not known if an individual recovered
and the testing procedures often do not depend on an earlier infection.
Note that $N(t)+I(t)=N(0)=S(0)$ is constant. 

While for each value of the risk, $I(t)$ intervenes as a multiplicative factor for the above conditional quantities, $I_r$ and $N_r$ do not necessarily evolve linearly with $I(t)$ without further assumptions due the dependency on $p(r|I(t),\sigma)$. We however make the following additional assumptions in line with the above section of Supplementary Methods on SIR modeling.
\begin{itemize}
\item $N(t)$ remains close to $N(0)$ for all times, i.e., $I(t)$ is 
always much smaller than $N(t)$ reflecting 
that only a small proportion of the population is  infected at any time,
	\item $I_r$ remains small in comparison to $N_r$ at any given score value $r$  (reflecting the low proportion of infectious within the population, and that the risk value achieved by an individual does not provide strong evidence for her being infectious), 
	\item the modulation by $I(t)$ of $p(r|I(t),\sigma)$ is small with respect to its marginal $p(r|\sigma)$, such that $p(r|I(t),\sigma)=p(r|\sigma)+\epsilon u(r,I(t),\sigma)$, for $\epsilon$ small and $|u(r,I(t),\sigma)|<p(r|\sigma)$
\end{itemize}

\subsubsection*{Specific testing regimes}
Inference on the growth rate $\lambda(t)$ based on equation~\ref{eq:firstOrdI} 
requires estimation of the true number of infectious $I(t)$. In practice only the time evolution
of confirmed cases $Y(t)$ is known but this is not necessarily a good indicator of $I(t)$ as it also depends on the amount of testing $T(t)$. Taking the above risk based model, we assume the pool of $T$ tested individuals is chosen as the $T$ having highest risk. This leads to the relation between $T$ and threshold risk $\rho(T)$ (taking the expectation)
\[
T = \int_{\rho(T)}^{\rmax}\left(N_r + I_r\right) dr\,.
\]

We assume moreover that  testing is "ideal" such that every test is 100\% reliable, performed only once per individual and instantaneous (putative lags are discussed in section \nameref{ssec:delays}). The expected number of confirmed cases is then
\[
\mathbb{E}[Y(t)] = \int_{\rho(T)}^{\rmax} I_r dr\,,
\]
with deviation from this expectation due to finite sampling (see section \nameref{ssec:poissonreg}).
Replacing with the above approximation we get
\[
\mathbb{E}[Y(t)] = I(t) \int_{\rho(T)}^{\rmax} p(r|\sigma =1) dr+ \epsilon I(t) \int_{\rho(T)}^{\rmax} u(r,I(t),\sigma =1) dr\,,
\]
which leads to, by using the assumptions on $u$, the following approximate expression in the form of a Taylor expansion:
\begin{equation}
\mathbb{E}[Y(t)] = I(t) \int_{\rho(T)}^{\rmax} p(r|\sigma =1) dr + o(I(t))
\end{equation}
implying that the testing models can be expressed as a function of the true number of infected and the number of tests
\begin{align}\label{eq:approxTestingIntegral}
\mathbb{E}(Y(t)) \approx I(t)f( T(t))\,\mbox{ with }\, f( T(t)) = \int_{\rho(T)}^{\rmax} p(r|\sigma =1) dr
\end{align}
and due to counting nature of the number of confirmed cases, we assume that $Y(t)$ follows a Poisson distribution.
We then consider the following testing regimes.

\paragraph{Adapted testing (baseline).} The baseline model assumes testing has no influence on observed cases
and $I$ and $Y$ are proportional
\begin{align}
\mathbb{E}[Y(t)] = \kappa I(t) = I(t)f_a(T(t)).
\end{align} 

This can be put in the above framework under the "adapted testing " assumption stating that: 
$T(t)$ is chosen in order to test all candidate individuals up to a fixed risk $r_0$, fixed by government response once and for all. Indeed, under the approximation of equation~\ref{eq:approxTestingIntegral}, this leads to
\begin{equation}\label{eq:adaptTestApprox}
\mathbb{E}[Y(t)] \approx I(t) \int_{r_0}^{\rmax} p(r|\sigma =1) dr\,,
\end{equation}
yielding the above linear relationship with $\kappa  = \int_{r_0}^{\rmax} p(r|\sigma =1) dr$, which does not depend on $I(t)$.
On the other hand, to understand how the amount of testing is modulated, we can solve for $T$
\[
\rho(T) = r_0\,
\] 
leading to
\begin{multline}
T(r_0)  = \int_{r_0}^{\rmax}\left(N_r + I_r\right) dr
 = N(t)\int_{r_0}^{\rmax}p(r|I(t),\sigma=0)dr+I(t)\int_{r_0}^{\rmax}p(r|I(t),\sigma=1)dr\\
\approx N(t)\int_{r_0}^{\rmax}p(r|\sigma=0)dr+\epsilon N(t) \int_{r_0}^{\rmax}u(r,I(t),\sigma=0)dr +I(t)\int_{r_0}^{\rmax}p(r|\sigma=1)dr+o(I(t))\\
\approx (N(0)-I(t))\int_{r_0}^{\rmax}p(r|\sigma=0)dr+\epsilon N(0) \int_{0}^{r_0}u(r,I(t),\sigma=0)dr +I(t)\int_{r_0}^{\rmax}p(r|\sigma=1)dr\,,\\
\approx N(0)\int_{r_0}^{\rmax}p(r|\sigma=0)dr+\epsilon N(0) \int_{r_0}^{\rmax}u(r,I(t),\sigma=0)dr +I(t)\int_{r_0}^{\rmax}\left(p(r|\sigma=1)-p(r|\sigma=0)\right) dr\,.
\end{multline}
In this last approximation, the first term is constant, while the next two terms may vary as a function of the prevalence $I(t)$, in principle with comparable magnitude and also at least comparable to the variations in the number of confirmed cases of equation~\ref{eq:adaptTestApprox}. Notably, variations in the second term, reflecting the number of non-infectious, may be caused by contact tracing measures, leading, when $I(t)$ gets large, to a larger number of non-infectious contacts assigned high values of $r$, and thus being tested. The third term trivially reflects the shift in risk-structure due to shifting from the non-infectious structure to the infectious structure as $I(t)$ increases.
This justifies that the adapted testing assumption entails a putative dependency of the number of test $T(t)$ on $I(t)$, justifying the causal graph of Fig.~\ref{fig:theory}B (top). 

\paragraph{Limiting factor testing.} 
Due to material or organizational limitations of the testing procedure, all individuals with risk above $r_0$ may not be tested. We model this regime by assuming that tests are performed following the risk of individuals in decreasing order, until the testing capacity, fixed to $T(t)$, is completely exploited. Contrary to the above example, $T(t)$ is now influencing $Y(t)$ through the relation
\[
\mathbb{E}[Y(t)] = I(t) \int_{\rho(T(t))}^{\rmax} p(r|I(t),\sigma =1) dr\,.
\]
We use additional assumptions to simplify this relation. Specifically, we assume 
a constant ratio 
\begin{equation}\label{eq:constRatioAssum}
p(r|I(t), \sigma=0)/p(r|I(t),\sigma=1)=\mu_0 \quad  \mbox{for } r\geq \rho(T(t))\quad \mbox{at all times.}
\end{equation}
This leads to 
\begin{align}\label{eq:test_relation}
\begin{split}
T(t) &= \int_{\rho(T)}^{\rmax}\left(N_r + I_r \right) dr = N(t)\int_{\rho(T)}^{\rmax} p(r|I(t), \sigma=0)dr + I(t)\int_{\rho(T)}^{\rmax} p(r|I(t), \sigma=1)dr\,,\\
 & = (N(0)-I(t))\mu_0\int_{\rho(T)}^{\rmax} p(r|I(t), \sigma=1) dr+ I(t)\int_{\rho(T)}^{\rmax} p(r|I(t), \sigma=1)dr\,,\\
 & = \left(\mu_0 N(0)+(1-\mu_0)I(t)\right)\int_{\rho(T)}^{\rmax} p(r|I(t), \sigma=1) dr\,.
 \end{split}
\end{align}
such that the $\rho(T(t))$ dependent term in the expectation of confirmed cases writes
\[
\int_{\rho(T(t))}^{\rmax} p(r|I(t),\sigma =1) dr=\frac{T(t)}{\mu_0N(0)+(1-\mu_0)I(t)}\approx \frac{T(t)}{\mu_0N(0)}+o(I(t))\, 
\]
leading to (neglecting the order two term in $I(t)$)
\[
\mathbb{E}[Y(t)] \approx \frac{1}{\mu_0N(0)}I(t) T(t)\,.
\]
Thus the expected number of cases takes the form of a mass action law, with proportionality to both the number of tests $T$ and to the number of infections $I$ that governs the probability that a randomly tested individual is infected, i.e., in our general testing model framework
\begin{align}
	\mathbb{E}(Y(t)) = \kappa I(t) T(t) = I(t)f_l(T(t)).
\end{align}
Given detecting a new case relies on both testing the subject and that the subject is infected, the above approximation can be interpreted as a law of mass action for testing, that induces a multiplicative effect
of the rate of new tests on the rate of new confirmed cases.

\paragraph{Up-saturating testing model.}
Based on a similar idea as in the previous model we in addition assume that with increasing
test numbers the probability of a test to be positive decreases, i.e.,
the quotient $I_r/N_r$ increases as the risk $r$ decreases. 
We relax the assumption that $p(r|\sigma=1)$ is constant and instead consider the following parametric form
\begin{align*}
p(r |I(t), \sigma=0) & = p(r | \sigma=0) = \nu_0\,,\\
p(r |I(t), \sigma=1) & = p(r | \sigma=1) = \nu_1 \frac{\omega_0^2}{(\rmax - r +\omega_0)^2}\,.
\end{align*} 
This expression is almost constant and equal to $\nu_1$ for $r> \rmax- \omega_0$ but then decays quickly accounting for
saturation of the testing policy.
Using manipulations similar to \eqref{eq:test_relation} 
we obtain the following expression for the risk threshold
\begin{align}\label{eq:risk_threshold}
\begin{split}
T(t) &= \int_{\rho(T)}^{\rmax}\left(N_r + I_r \right) dr \approx N(t)\int_{\rho(T)}^{\rmax} p(r|I(t), \sigma=0)dr
\approx N(0)\nu_0 (\rmax - \rho(T))\\
 \Rightarrow \rho(T)&=\rmax-\frac{T(t)}{\nu_0 N(0)}.
 \end{split}
\end{align}
The expected number of confirmed cases then reads
\begin{align*}
\mathbb{E}[Y(t)] &\approx I(t) \int_{\rmax -\frac{T(t)}{\nu_0 N(0)}}^{\rmax} p(r|\sigma =1) dr\,,
=  I(t) \nu_1 \left[ \frac{\omega_0^2}{\rmax - r + \omega_0}\right]^{r=\rmax}_{r=\rmax -T(t)/(N(0)\nu_0)}
\\
&=
I(t)   \nu_1\omega_0\frac{T(t)/(N(0)\nu_0)}{T(t)/(N(0)\nu_0) + \omega_0}.
\end{align*}
Upon reparametrizing $\kappa=\nu_1\omega_0$ and $\alpha=N(0)\nu_0\omega_0$ we
obtain the following testing model
\begin{align}
\mathbb{E}(Y(t)) = \kappa I(t) \frac{T(t)}{T(t)+\alpha} = I(t) f_u( T(t)).
\end{align}
Note that for small numbers of tests (i.e., $T(t)<\alpha$) this model behaves similarly to the limiting factor testing model
$f_l$, i.e., every test has the same probability to be positive 
and confirmed cases are proportional to $I(t)T(t)$
while for large numbers of tests 
$Y(t)\approx \kappa I(t)$ and the model approaches the baseline $f_a$.
Thus this model interpolates between the two regimes.

We define the upper threshold of this model as $T(t)=\alpha$, the point where the prediction changed by a factor of 2 compared to the limiting factor regime approximation for small $T$, $Y(t) \approx \kappa I(t) T(t)/\alpha$.

\paragraph{Down-saturating testing model.}
We investigate one further model that accounts for an increasing ratio $I_r/N_r$.
The idea of this testing model is that 
 a certain fraction of strongly symptomatic patients and very close contacts are always discovered
 almost independently of the testing numbers. 
 In our framework this can be formalized by assuming that the distribution $p(r|\sigma=1)$
contains a point mass for $r=\rmax$, i.e., $\mathbb{P}(r=\rmax|\sigma=1)=\delta$.
In addition we assume that $p(r|\sigma=1)=\nu_1$ and $p(r|I(t),\sigma=0)=\nu_0$ as a special case of the limiting factor testing model. Based on  similar calculations as before
we  obtain
\begin{align*}
Y(t) = \delta I(t) + \kappa I(t) T(t) =  \kappa I(t) (\alpha + T(t)) = I(t) f_d(T(t)).
\end{align*}
For a justification of the names of the testing models we refer to the plot in 
Figure~\ref{fig:validModel}A.

We define the lower threshold of this model as $T(t)=\alpha$, the point where the prediction changed by a factor of 2 compared to the limiting factor regime approximation for large $T$, $Y(t) \approx \kappa I(t) T(t)$.

%
%

\section*{Validation of testing models through death predictions}
Death statistics of the Covid 19 pandemic are often assumed to be reported more accurately than the number of infections.
Therefore they can be used to 
check the validity and compare different testing models. In this section we describe the methodology underlying our validation approach.

\subsubsection*{Relating infections and deaths}\label{ssec:infectToDeaths}
We modeled the relation between the number of infections $I(t)$ and the number of deaths $D(t)$ following \cite{flaxmanNature}.
We assume that the expectation of the death rate $D(t)$  
 relates to the 
time series of daily incidence $i(t)$ through the equation
\begin{align}\label{eq:death_infection_relation}
\mathbb{E}(D) = \text{ifr} \cdot( \Pi \ast i)
\end{align}
where $\Pi$ denotes a filter representing the distribution of the time between infection and death, called \textit{onset-to-death distribution}, $\text{ifr}$
the infection fatality rate and $\ast$ the discrete convolution. 
We use the parametric model of onset-to-death distribution in days,
\begin{equation}
	\Pi' = .5\text{ Gamma}(k=4.39, \theta=1.16) + .5\text{ Gamma}(k=8.46, \theta=2.22)
\end{equation}
provided by \cite{flaxmanNature} based on the experimental results of \cite{verity2020estimates},
and discretize it to obtain one point per day, using $\Pi(t) = \mathbb{P}(\Pi'\in [t-0.5, t + 0.5])$.
We will assume that $\text{ifr}$ remains constant in time but may differ between countries. The actual value is not relevant to our analysis.

\subsubsection*{Validation of testing models}\label{ssec:validTest}
We can now combine the relation between infections and deaths and the testing
model and assess how well different testing models can explain the observed 
evolution of cases, deaths, and tests.
Recall that we assume 
\begin{align}\label{eq:test_model_recap}
\mathbb{E}(Y(t)) = I(t) f( T(t))
\end{align}
for some testing model $f$.
We can infer
an unbiased estimate of $I$ as
\begin{align}\label{eq:test_model_inverse}
\hat{I}(t) = \frac{Y(t)}{ f(T(t))}
\end{align}

Since our testing model involves the prevalence while the expression for the expected
mortality rate involves the incidence we need to connect incidence and prevalence in the SIR model.
In the SIR model the incidence agrees with the gain term $i(t)=S(0)\beta(t)I(t)$.
Discretization on a daily level leads to the expression 
 $I(t) - I(t-1) \approx I'(t)= (S(0)\beta(t-1) - \gamma) I(t-1)$ where
$\gamma$ is the recovery rate in the SIR model. Combining the last to equations implies
\begin{align}\label{eq:new_infections_from_total}
i(t) = I(t) - (1 - \gamma) I(t-1).
\end{align} 
 The recovery rate $\gamma$ in the SIR model corresponds to the inverse of the mean 
 generation interval which we assume to be 5.0 days based on \cite{Ferrettieabb6936}, 
 thus $\gamma =0.2\;  \text{days}$.
Our results are not sensitive to the value of $\gamma$ and we can also use the prevalence $I$ as a proxy for the incidence $i(t)$. 
 We use testing time series $T(t)=T^C(t)$ and confirmed cases $Y(t)=Y^C(t)$ from some country $C$ to estimate 
 infection numbers using \eqref{eq:test_model_inverse}, i.e., $\hat{I}(t)= Y(t) / f_m( T(t))$.
 Then we infer an estimate $D_{est}^C(t) \propto (\Pi \ast \hat{i})(t) $
for the expected number of deaths
 using the relations \eqref{eq:death_infection_relation} and \eqref{eq:new_infections_from_total}.
 
 We assess the quality of a testing model by its ability to predict the observed time series
 of deaths. Since in most countries the reported number of deaths drops sharply at weekends we replace
 the reported number of deaths by a 7 day rolling average $D^C$.
We measure the distance between $D^C$ and $D^{C}_{est}$ by
\begin{align*}
\Delta(C,f_m)= \mathrm{Var}\left[\ln(D^C(t)) - \ln(D^{C}_{est}(t))\right]
\end{align*} 
where the time series is restricted to days with at least 10
observed deaths. We remark that the variance does not depend
on the unspecified constant of proportionality given by $ifr$.
We average this across countries using the expression
\begin{align*}\label{eq:prediction_error}
\Delta_{av}(f_m) =\frac{1}{\# \,\text{countries}} \sum_{C} \frac{\Delta(C, f_m)}{\mathrm{Var}(\ln(D^C)) + 1}.
\end{align*}
The normalisation by the variance of $\ln(D^C)$ ensures  that 
the results of different countries are comparable and the error is not dominated by 
countries with bad data quality.
\subsection*{Optimisation of testing models}\label{ssec:optim}
Here we describe how the parameters of the testing models are optimized such that the
prediction error of equation~\ref{eq:prediction_error} is minimized.


Our error measure is insensitive to multiplicative factors such that we do not need 
to optimize the factor $\kappa$ contained in all testing models.
This allows the testing models to adapt for different infection fatality
rates due to varying age distributions and health systems.

The further parameters involved in the testing model are assumed to be independent of the country
and optimized using
10-fold
cross validation and automatic differentiation and the BFGS optimizer using the pyTorch library \cite{NEURIPS2019_9015}.

We also allow the parameter $\alpha$ of the testing model to be a linear function of a set of $n$ covariates $X_i$ of
the countries, i.e., 
\begin{align}
\alpha = \beta_0 + \sum_{i=1}^n \beta_i X_i
\end{align}
which can be optimized for $n+1$ dimensional vector $\beta$.
%
\section*{Estimation of causal effects}\label{ssec:causEffect}
Suitable testing model provide estimates for the true dynamics of infection $I(t)$
and therefore allow to infer changes  of $\lambda(t)$, which controls the growth or decay of $I(t)$.
The growth rate $\lambda(t)$
crucially depends (through the transmission rate $\beta(t)$) on the current social distancing policy
$L(t)$, which affects the contact between individuals in the population.
For simplicity we assume a binary social distancing policy $L(t)=H(t-t_L)$ 
where  $H$ is the Heaviside function (such that $H(t)=1$ if $t\geq 0$ and $H(t)=0$ otherwise) and
$t_{L}$  is the time where the lockdown begins.
This results in a step function effect on the growth rate
\[
\lambda(t)=\lambda_{0}-\theta_{0}H(t-t_L)=\beta_{0}S(0)-\gamma-\theta_{0}H(t-t_L)\,,
\]
with   $\lambda_{0}$ the baseline value of $\lambda(t)$ under normal conditions
(no social distancing), related to $\beta_0$ the baseline disease transmission rate, and $\theta_{0}$ is the causal effect of social distancing. As a consequence, the logarithmic number of infected evolves in time as
\begin{equation}\label{eq:growth_I}
 \log I(t)=\log I(0)+t \lambda_0 - (t-t_L)\theta_0 H(t-t_L)
\end{equation}
leading to a piecewise linear time course as illustrated on Fig.~\ref{fig:theory}D, 
where the change in slope reflects directly the causal effect of social distancing.

We now include the testing model relation between $I$, $Y$, and $T$ 
in thea piecewise linear evolution of  $\log I(t)$ as in \eqref{eq:growth_I}
and we obtain
\begin{align}\label{eq:expansion_exposure}
\log \mathbb{E}[Y (t)]=\log I(t) + \log f(T(t))=\log I(0)+\lambda_0 t -(t-t_L) \theta_0 H(t-t_L)+\log f(T(t))
\end{align}
The evolution of $\log \mathbb{E}(Y(t))$ is thus guaranteed to reflect the piece-wise
linear trajectory of $I(t)$ only if the correction accounting for varying 
testing is included in the model.


\subsection*{Fitting data with Poisson regression}\label{ssec:poissonreg}

Given the log-linear form of the testing models of the theoretical value of $Y(t)$ (based on equation~\ref{eq:testingModelGeneric}, an appropriate statistical framework for estimating $\lambda(t)$ is Poisson regression \cite{peng2006model}, which is a particular form of Generalized Linear Model. In this setting, observations of the number of confirmed cases $\widehat{Y}(t)$ are assumed Poisson distributed, with expectation parameter $\overline{Y}(t) = \mathbb{E} [Y(t)]$ modeled with intercept $a$ and slope $b$ as
\begin{equation}\label{eq:regUncorrected}
\log \mathbb{E} [{Y}(t)] = a + t b\,,
\end{equation}
in the uncorrected case (adapted testing), such that the regression coefficient $b$, which is the slope of the curve schematized in Fig.~\ref{fig:theory}D,
 provides an estimate of $\lambda(t)$ on time intervals where it is assumed constant. The model dependent correction for the varying testing rates can also be incorporated as in \eqref{eq:expansion_exposure} in the Poisson regression as an \textit{exposure} term, leading to the corrected model
\begin{equation}\label{eq:regCorrected}
\log \mathbb{E} [{Y}(t)] = a + t b  +\log f(T(t))\,,
\end{equation}
with $f$ the function associated to the testing model ($f$ is identity for the limiting model, and constant for the adapted model).
These statistical models allow inferring the disease dynamics from observational data.


\newcommand{\validationnrcountries}{66}
\newcommand{\validationmindeath}{1000}
\newcommand{\validationmincases}{20}
\newcommand{\validationcovnrcountries}{62}

\newcommand{\daysbeforeld}{7}
\newcommand{\ldmincases}{10}

\newcommand{\daysafterld}{5}
\newcommand{\daysafterldrestart}{10}

\newcommand{\lastday}{2020-11-22}
\newcommand{\nrcountsecond}{13}
\newcommand{\nrcountfirst}{13}

\subsection*{Data preprocessing}\label{ssec:preproc}
We use the Our World in Data Covid 19 dataset available online at 
\href{ourworldindata.org}{ourworldindata.org} and published in
\cite{Hasell2020}. This dataset contains data for over 200 entities
and provides various time resolved data related to Covid 19. 
Our analysis relies on the number of confirmed cases, testing information and fatality numbers.

\subsubsection*{Reporting delays} \label{ssec:delays}
The available data typically  contains a time lag $\delta$ 
 between the date of a test and the date the result of this test is reported. 
 This implies that the testing models actually relate
 test statistics of time $t$ with the prevalence $I(t-\delta)$ at an earlier point. 
 We account for this in the analysis of causal effects by excluding some days from the regression (see below). In the validation analysis we assume that for each country the reporting delay  for fatalities and tests is similar such that the offsets cancel.
Moreover, this delay is small in regard to the time scale of the death rate trajectory.

\subsubsection*{Exclusion-selection of countries for the death prediction} \label{ssec:excl_sel}
For the model validation based on the death prediction we use all countries for which 
at least weekly testing information is available and which 
had at least \validationmindeath \ fatalities attributed to Sars-Cov 2 such that death prediction is meaningful. In addition we removed China from the data because testing started in a late stage of the epidemics there. This left \validationnrcountries  \  countries for the data analysis.

We also investigated the dependence of the testing model on additional country covariates.
The covariate data was taken from the DELVE Global COVID-19 Dataset\footnote{\url{github.com/rs-delve/covid19_datasets}} \cite{DelveCovidDataset} and was originally collected
by the World Bank. 
Data was available for \validationcovnrcountries \ of the previously selected countries.
We started the time series when at least \validationmincases \ cases were reported because some of the testing models notably the limiting factor model become
unstable for very
small case numbers.

\subsubsection*{Interpolation of missing data}
We base our analysis on reports of daily new tests and confirmed cases. In case daily updates are missing for n days, we perform linear interpolation of the logarithm of the cumulative number of test and cases, respectively, and use the daily difference to approximate the daily updates. In this way, the cumulative number remains consistent with the observations, while the daily updates are interpolated.

\subsubsection*{Choice of lockdown dates and country selection} \label{ssec:lockdown}

For the analysis of lockdown effects we also relied on the Our World in Data dataset.
The lockdown dates for the spring period was determined based on information reported in the BBC article \href{https://www.bbc.com/news/world-52103747}{\textit{Coronavirus: The world in lockdown in maps and charts}}\footnote{\url{https://www.bbc.com/news/world-52103747}} where the earliest date among ``national recommendation'' and ``national lockdown'' was chosen.

From those countries we selected all countries that had at least \ldmincases \ cases reported \daysbeforeld \ days
before the beginning of the lockdown and at least weekly testing information.
The \nrcountfirst \ selected countries and lockdown dates can be found in 
Table~\ref{tab:ld}.

For the pre-lockdown Poisson regressions  we considered the time interval spanning from the first day
with at least \ldmincases \ cases until \daysafterld \ days after the onset of the lockdown taking into account that due to the incubation period and delays in case reporting this time interval 
captures infections before the lockdown. 
For the Poisson regressions  during the lockdown we considered the interval starting \daysafterldrestart
\ days after the lockdown onset and ending when
the mobility reduction was less than 80\% of the maximal reduction in mobility as measured by 
the Google mobility reports\footnote{available at \href{google.com/covid19/mobility}{google.com/covid19/mobility/}} where we used weekly averages of the sum of retail, transit stations, and workplace indicators.

Many, mostly European, countries  ordered a second lockdown in the fall of 2020. As the measures were installed more gradually the transition between pre-lockdown and lockdown was less sharp.
Therefore we determined the onset of the second lockdown based on the stringency index from the Oxford COVID-19 Government Response Tracker (OxCGRT)
\cite{stringency} which was 
designed to measure the stringency of the governmental responses to the Covid-19 pandemic on a scale from 0-100.
We defined the beginning of the lockdown as the first day where the stringency index was above 50.
The data was taken from the Our World in Data dataset.
Using this definition of the second lockdown \nrcountsecond \ countries issued a second lockdown
(some countries were excluded because they never lifted the stringency of their measures below 50 and
Thailand was excluded because there was not enough testing data available for the post lockdown period)
and their lockdown dates can be found in Table~\ref{tab:ld2}.

We remark that while the definitions of the two lockdowns do not agree using 
the stringency based definition would only change the lockdown dates of the first lockdown by a few days.

For the pre-lockdown regressions for the estimation of causal effects we used the 3 weeks before the lockdown and for the regressions during lockdown we used data from the 3 weeks starting 10 days after the onset of the lockdown to again account for reporting delays.

\subsubsection*{Data selection for the US}\label{ssec:usselect}
We performed a similar analysis for the states of the USA based on data provided by the
COVID Tracking project\footnote{available at \href{covidtracking.com}{covidtracking.com}}. 
The lockdown dates for the US states were chosen as the day of the 'stay at home order'.
Dates were taken from from Wikipedia\footnote{available at 
\url{https://en.wikipedia.org/wiki/U.S._state_and_local_government_responses_to_the_COVID-19_pandemic}}.
There were 30 states that had sufficient data for the lockdown analysis. They can be found along with their lockdown dates in Table~\ref{tab:ld3}.
For the validation analysis we used the same criteria as above which were satisfied by 40 states.
We did not pursue the addition of covariates for the parametric testing models.

%

\bibliographystyle{Science}
\renewcommand\refname{References and notes} 

\bibliography{covid}

\begin{thebibliography}{10}

\bibitem{Giordano2020}
G.~Giordano, {\it et~al.\/}, {\it Nature Medicine\/} {\bf 26}, 855 (2020).

\bibitem{Dehningeabb9789}
J.~Dehning, {\it et~al.\/}, {\it Science\/}  (2020).

\bibitem{hart2020mediacoverage}
P.~S. Hart, S.~Chinn, S.~Soroka, {\it Science Communication\/} {\bf 42}, 679
  (2020).

\bibitem{Tejedor2020}
S.~Tejedor, L.~Cervi, F.~Tusa, M.~Portales, M.~Zabotina, {\it International
  journal of environmental research and public health\/} {\bf 17}, 6330 (2020).
  32878092[pmid].

\bibitem{Cinelli2020}
M.~Cinelli, {\it et~al.\/}, {\it Scientific Reports\/} {\bf 10}, 16598 (2020).

\bibitem{kermack1932contributions}
W.~O. Kermack, A.~G. McKendrick, {\it Proceedings of the Royal Society of
  London. Series A\/} {\bf 138}, 55 (1932).

\bibitem{kermack1933contributions}
W.~O. Kermack, A.~G. McKendrick, {\it Proceedings of the Royal Society of
  London. Series A\/} {\bf 141}, 94 (1933).

\bibitem{m1925applications}
A.~M'Kendrick, {\it Proceedings of the Edinburgh Mathematical Society\/} {\bf
  44}, 98 (1925).

\bibitem{causality}
J.~Pearl, {\it Causality: Models, Reasoning and Inference\/} (Cambridge
  University Press, USA, 2009), second edn.

\bibitem{elementsofcausal}
J.~Peters, D.~Janzing, B.~Sch{\"o}lkopf, {\it Elements of Causal Inference -
  Foundations and Learning Algorithms\/}, Adaptive Computation and Machine
  Learning Series (The MIT Press, Cambridge, MA, USA, 2017).

\bibitem{Hasell2020}
J.~Hasell, {\it et~al.\/}, {\it Scientific Data\/} {\bf 7}, 345 (2020).

\bibitem{verity2020estimates}
R.~Verity, {\it et~al.\/}, {\it The Lancet infectious diseases\/}  (2020).

\bibitem{flaxmanNature}
S.~Flaxman, {\it et~al.\/}, {\it Nature\/} {\bf 584}, 257 (2020).

\bibitem{NEURIPS2019_9015}
A.~Paszke, {\it et~al.\/}, {\it Advances in Neural Information Processing
  Systems 32\/}, H.~Wallach, {\it et~al.\/}, eds. (Curran Associates, Inc.,
  2019), pp. 8024--8035.

\bibitem{doi:10.1098/rspb.2006.3754}
J.~Wallinga, M.~Lipsitch, {\it Proceedings of the Royal Society B: Biological
  Sciences\/} {\bf 274}, 599 (2007).

\bibitem{10.1093/jtm/taaa021}
Y.~Liu, A.~A. Gayle, A.~Wilder-Smith, J.~Rocklöv, {\it Journal of Travel
  Medicine\/} {\bf 27} (2020).

\bibitem{Saljeeabc3517}
H.~Salje, {\it et~al.\/}, {\it Science\/}  (2020).

\bibitem{imai2020report}
N.~Imai, {\it et~al.\/}, {\it Imperial College London\/}  (2020).

\bibitem{CHERNOZHUKOV2020}
V.~Chernozhukov, H.~Kasahara, P.~Schrimpf, {\it Journal of Econometrics\/}
  (2020).

\bibitem{contreras2020longterm}
S.~Contreras, J.~Dehning, S.~B. Mohr, F.~P. Spitzner, V.~Priesemann, Towards a
  long-term control of covid-19 at low case numbers (2020).

\bibitem{rtlive2020}
K.~Systrom, T.~Vladek, M.~Krieger, Rt.live,
  \url{https://github.com/rtcovidlive/covid-model} (2020).

\bibitem{karin2020adaptive}
O.~Karin, {\it et~al.\/}, {\it medRxiv\/}  (2020).

\bibitem{bassi2020observed}
F.~Bassi, G.~Arbia, P.~Falorsi, {\it Science of The Total Environment\/} p.
  142799 (2020).

\bibitem{rostami2020sars}
A.~Rostami, {\it et~al.\/}, {\it Clinical Microbiology and Infection\/}
  (2020).

\bibitem{van2017reproduction}
P.~van~den Driessche, {\it Infectious Disease Modelling\/} {\bf 2}, 288 (2017).

\bibitem{webb1985theory}
G.~F. Webb, {\it et~al.\/}, {\it Theory of nonlinear age-dependent population
  dynamics\/} (CRC Press, 1985).

\bibitem{Ferrettieabb6936}
L.~Ferretti, {\it et~al.\/}, {\it Science\/} {\bf 368} (2020).

\bibitem{peng2006model}
R.~D. Peng, F.~Dominici, T.~A. Louis, {\it Journal of the Royal Statistical
  Society: Series A (Statistics in Society)\/} {\bf 169}, 179 (2006).

\bibitem{DelveCovidDataset}
Delve global covid-19 dataset,
  \url{https://github.com/rs-delve/covid19_datasets/blob/master/dataset/combined_dataset_latest.csv}.
  Accessed: 01/28/21.

\bibitem{stringency}
T.~Hale, {\it et~al.\/}, {\it Blavatnik School of Government Working Paper\/}
  (2020).

\end{thebibliography}

\section*{Acknowledgments}
\textbf{Authors contributions:} Conceptualization: M.B., S.B. and B.S.; Methodology: M.B., S.B. and B.S.; Software: M.B. and S.B.; Validation: M.B. and S.B; Formal analysis: M.B. and S.B.; Writing - original draft preparation: M.B. and S.B.; Writing - review and editing: M.B., S.B. and B.S.; Visualization: M.B. and S.B.; Supervision: M.B. and B.S. \textbf{Competing interests:} All authors declare no competing interests. \textbf{Data and materials availability:} All data will be available in the manuscript, the Supplementary Materials or online upon acceptance.

\newpage

\part*{}
\makeatletter
\begin{center}
{\Large Supplementary Information for\\}
{\Large \itshape\@title }\\
{ \large	{Michel Besserve, Simon Buchholz and Bernhard Sch\"olkopf}}\\
{Correspondence to : michel.besserve@tuebingen.mpg.de}
\end{center}
\makeatother

\textbf{This PDF file includes:}
\begin{itemize}
	\item Supplementary Figs. 1 to 4
	\item Tables S1 to S3
\end{itemize}
\clearpage

\renewcommand{\figurename}{\textbf{Supplementary Figure}}  
\renewcommand{\tablename}{\textbf{Supplementary Table}} 
\setcounter{figure}{0}      
\setcounter{table}{0}

\begin{figure}
	\includegraphics[width=\linewidth]{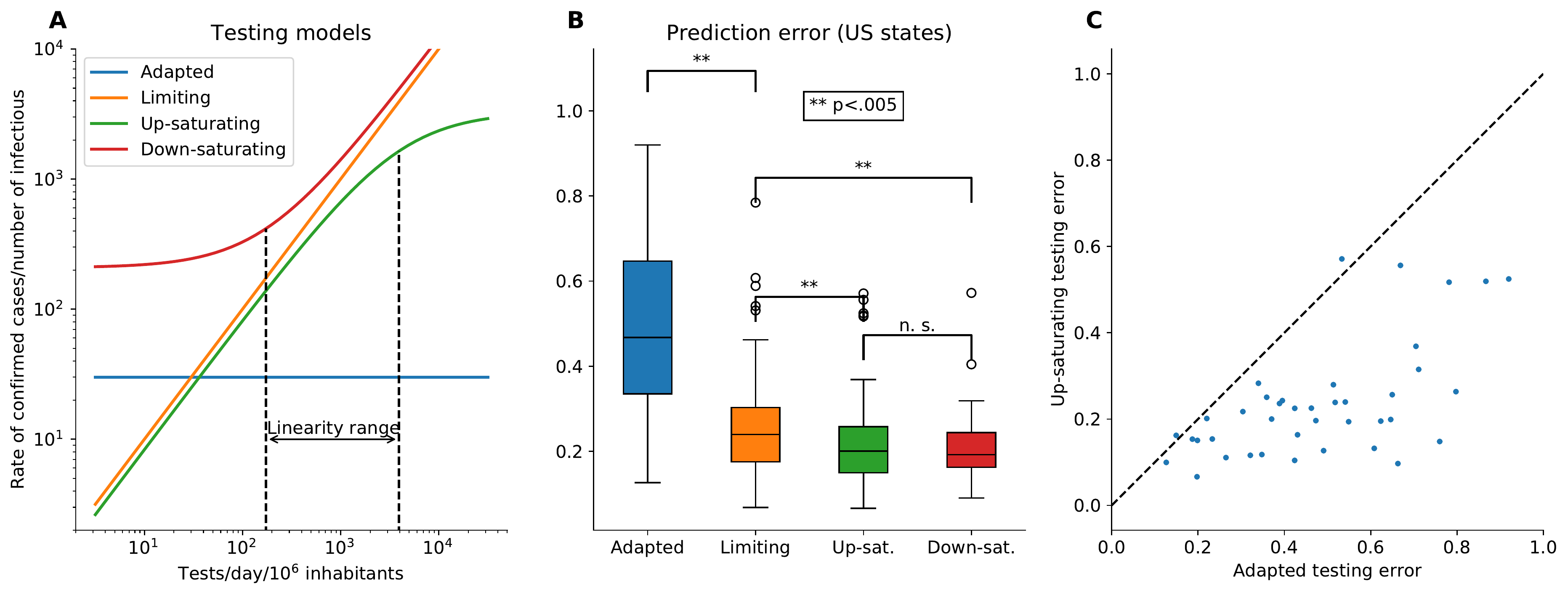}
	\caption{\textbf{Testing model validation for US states, related to Fig.~\ref{fig:validModel} } \textbf{(A-C)} Same as Fig~\ref{fig:validModel}B-D for the US states dataset, i.e.:
	\textbf{(A)} Functions associated with different testing models, with illustrative choice of parameters.
	\textbf{(B)} State-wise cross-validation prediction error for  death rate using different models linking these quantities to incidence. Significance was assessed using Wilcoxon signed rank tests. \textbf{(C)} Scatter plot showing the superiority of the limiting factor testing model, relative to the adapted testing model.	
	\label{sfig:modelval}}
\end{figure}

\clearpage

\begin{figure}
	\includegraphics[width=\linewidth]{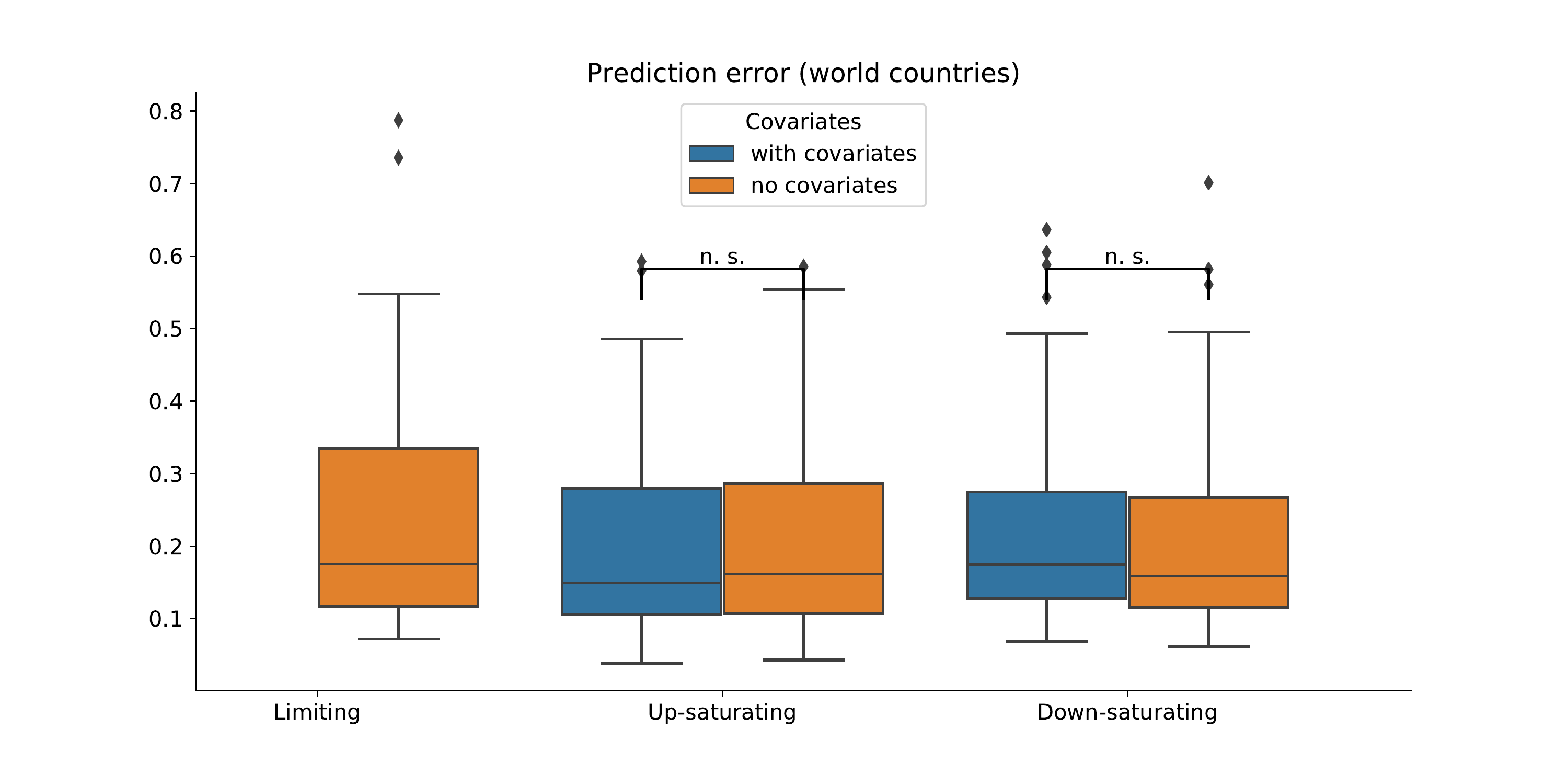}
	\caption{\textbf{Comparison of testing models with and without covariates, related to Fig.~\ref{fig:validModel}.}
	Country wise cross validation prediction error for the death rate for different testing models.
	Blue box-plots indicate regression without covariates.
	For orange box plots, the testing models' parameter was optimized using a linear dependence with respect to the country's Gross Domestic Product (GDP) per capita, population density, and the percentage of urban population as covariates. 
Note that both boxplots agree for the case of the limiting testing model which has no parameters to optimize. Statistical significance was assessed using a paired Wilcoxon signed rank test ($N=\Nworldvalidcovariates$). 	
		 \label{sfig:covariates}}
\end{figure}

\clearpage

\begin{figure}
	\centering
	\includegraphics[width=.5\linewidth]{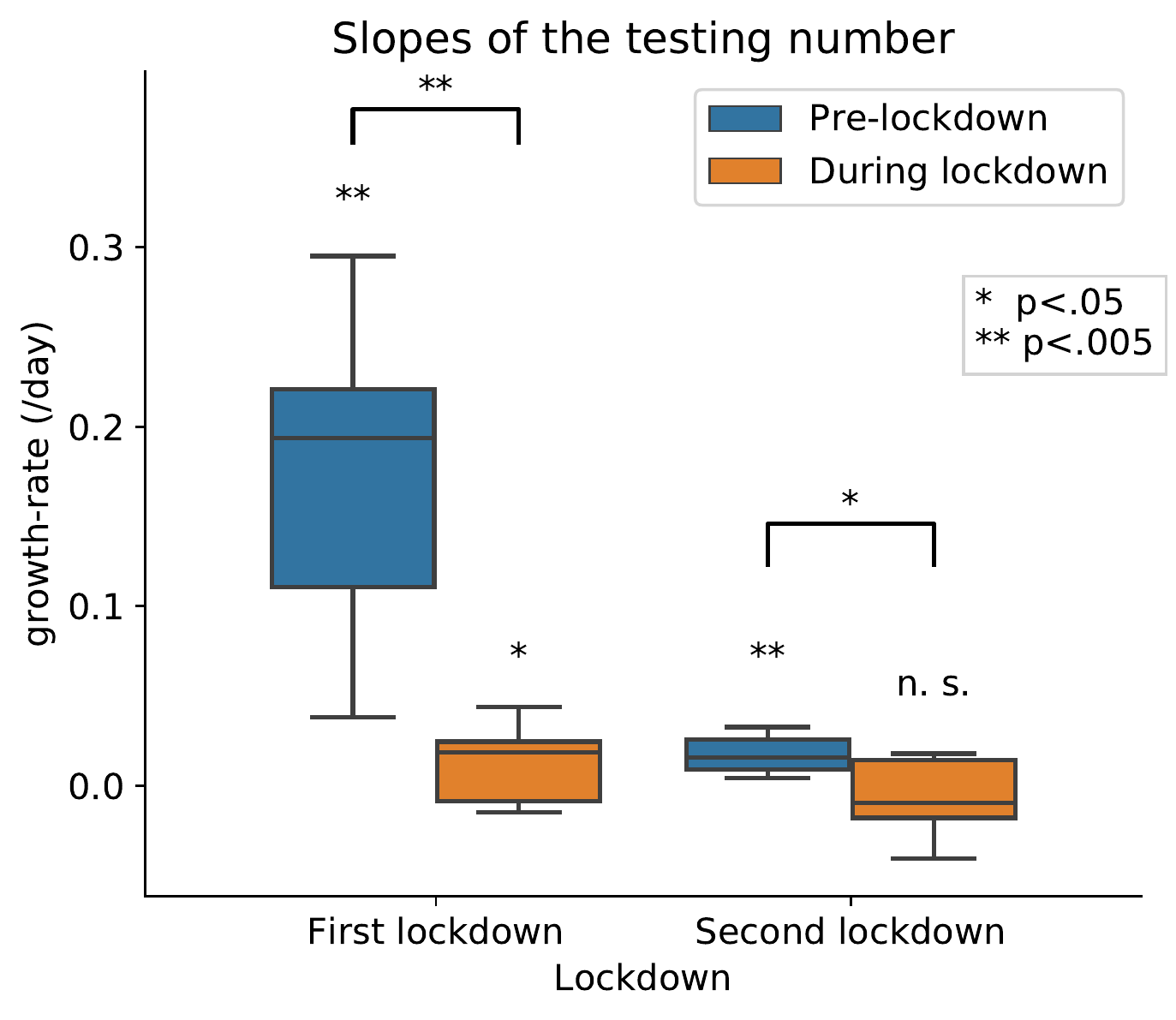}
	\caption{\textbf{Testing statistics, related to Fig.~\ref{fig:policyeval}.}
		Statistics of the exponential growth rate of the  number of tests estimated by Poisson regression for each
			country individually, separately for time periods before and during application of social distancing measures for the first and second lockdown.
	Significance was assess using Wilcoxon signed rank tests on the sign of the median of the regression coefficients or their difference (when above a line connecting two quantities). 
	\label{sfig:policyeval}}
\end{figure}

\clearpage

\begin{figure}
	\vspace*{-1cm}
	\includegraphics[width=\linewidth]{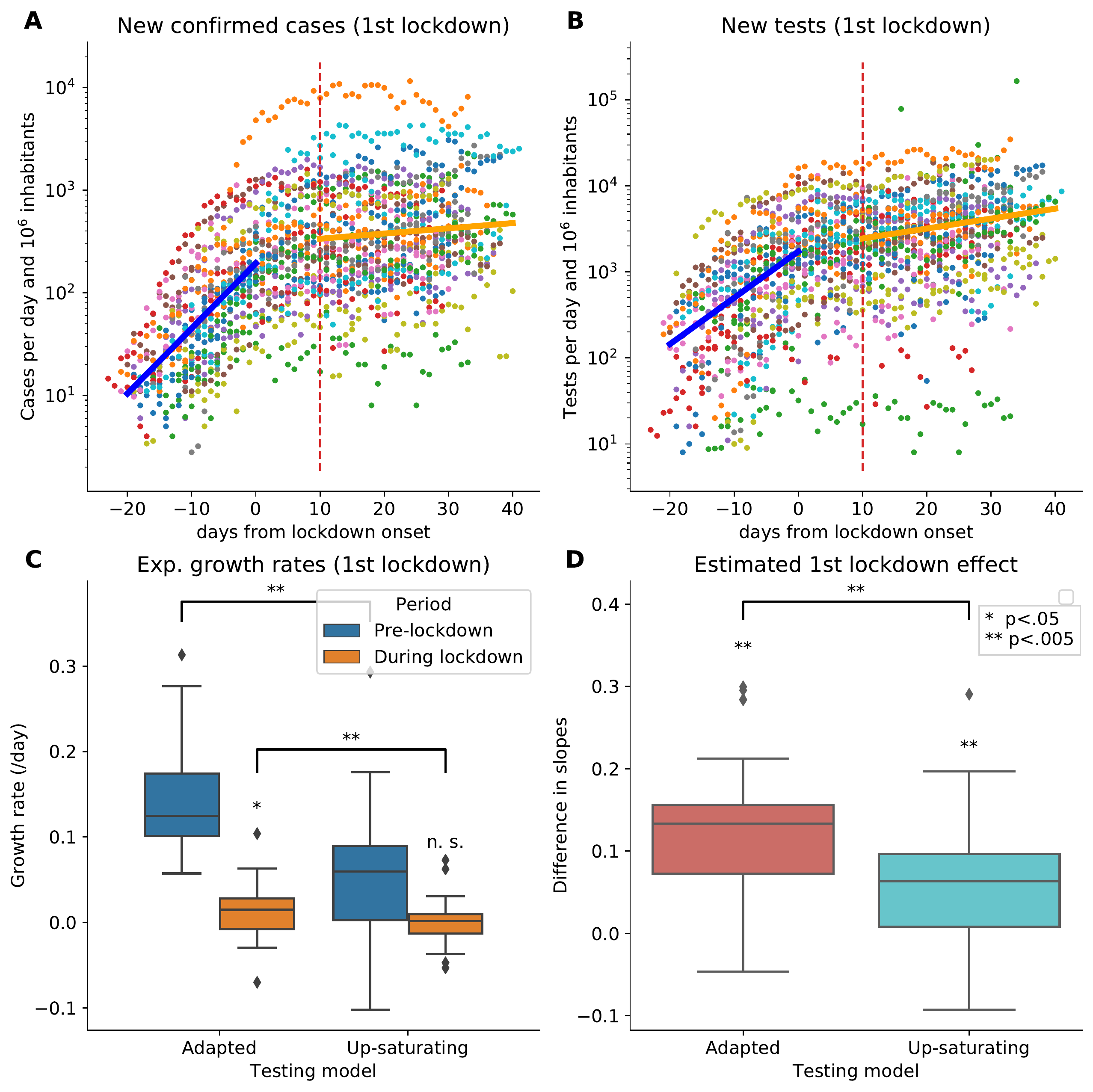}
		\caption{\textbf{Impact of testing assumptions on growth rate and policy evaluation for US states, related to Fig.~\ref{fig:policyeval}.} \textbf{(A)} Number of
			daily new confirmed cases for each country of the first lockdown. Blue and orange solid lines represent the average of Poisson regression models across countries, before and during social distancing, respectively. \textbf{(B)} Same as A for daily new tests
			per thousands (right). \textbf{(C)} Statistics of the exponential growth rate of the  prevalence estimated by Poisson regression for each
			country individually, separately for time periods before and during application of social distancing measures for the first lockdown. Testing is corrected for according to two testing models: adapted testing (left) and limited testing (right). 
			\textbf{(D)} Estimate of the effect of social distancing on transmission rate, with or without correction for testing. Indicated p-values correspond to Wilcoxon signed tests on the sign of the median of the regression coefficients or their difference (when above a line connecting two quantities). 		
		\label{sfig:USpolicyeval}}
\end{figure}

\clearpage

\begin{table}
	\caption{Lockdown dates and countries selected for the analysis of the first lockdowns, related to Fig.~\ref{fig:policyeval}. \label{tab:ld}}
	\centering
	\begin{tabular}{ l  r | l r}
		Country & Spring lockdown date & Country & Spring lockdown date
		\\  
		\hline			
Austria & 2020-03-13
&
Belgium & 2020-03-12
\\
Canada & 2020-03-13
&
Germany & 2020-03-18
\\
Italy & 2020-03-12
&
Malaysia & 2020-03-13
\\
Norway & 2020-03-12
&
Pakistan & 2020-03-23
\\
Switzerland & 2020-03-13
&
Thailand & 2020-03-20
\\
United Arab Emirates & 2020-03-26
&
United Kingdom & 2020-03-22
\\
United States & 2020-03-16
	\end{tabular}
\end{table}

\clearpage

\begin{table}
	\caption{Lockdown dates and countries selected for the analysis of the fall lockdowns, related to Fig.~\ref{fig:policyeval}. \label{tab:ld2}}
	\centering
	\begin{tabular}{ l  r | l r}
		Country & Fall lockdown date & Country & Fall lockdown date
		\\  
		\hline			
	Austria & 2020-10-17
&
Belgium & 2020-10-19
\\
Croatia & 2020-12-14
&
Finland & 2020-12-07
\\
France & 2020-10-30
&
Germany & 2020-10-15
\\
Hungary & 2020-11-05
&
Italy & 2020-10-06
\\
Lithuania & 2020-10-28
&
Luxembourg & 2020-10-20
\\
Norway & 2020-11-05
&
Romania & 2020-10-20
\\
Switzerland & 2020-11-02
	\end{tabular}
	\end{table}
	
	\clearpage

\begin{table}
	\caption{Lockdown dates and US states selected for the analysis related to Fig.~\ref{sfig:USpolicyeval}. \label{tab:ld3}}
	\centering
	\begin{tabular}{ l  r | l r}
		State & Lockdown date & State & Lockdown date
		\\  
		\hline			
AL & 2020-04-04
&
AZ & 2020-03-31
\\
CA & 2020-03-19
&
CO & 2020-03-26
\\
DC & 2020-04-01
&
FL & 2020-04-03
\\
GA & 2020-04-03
&
IL & 2020-03-21
\\
KS & 2020-03-30
&
LA & 2020-03-23
\\
MA & 2020-03-24
&
MD & 2020-03-30
\\
ME & 2020-04-02
&
MI & 2020-03-24
\\
MN & 2020-03-28
&
MO & 2020-04-06
\\
MS & 2020-04-03
&
NC & 2020-03-30
\\
NH & 2020-03-28
&
NJ & 2020-03-21
\\
NV & 2020-04-01
&
NY & 2020-03-22
\\
OH & 2020-03-24
&
OK & 2020-04-07
\\
PA & 2020-04-01
&
RI & 2020-03-28
\\
SC & 2020-04-07
&
TN & 2020-04-01
\\
TX & 2020-04-02
&
UT & 2020-04-02
\\
VA & 2020-03-30
&
WA & 2020-03-23
\\
WI & 2020-03-25
\end{tabular}
\end{table}

\end{document}